\documentclass{nature}
\usepackage{lineno}
\let\oldequation\equation
\let\oldendequation\endequation
\renewenvironment{equation}{\linenomathNonumbers\oldequation}{\oldendequation\endlinenomath}
\usepackage{graphicx}  
\usepackage{physics}  
\newcommand{\EXTTAB}[1] {Extended Data Table~\ref{#1}}
\newcommand{\EXTFIG}[1] {Extended Data Fig.~\ref{#1}}
\newcommand{\FIG}[1] {Fig.~\ref{#1}}
\newcommand{\TAB}[1] {Table~\ref{#1}}

\makeatletter
\let\saved@includegraphics\includegraphics
\AtBeginDocument{\let\includegraphics\saved@includegraphics}
\renewenvironment*{figure}{\@float{figure}}{\end@float}
\makeatother

\usepackage{color}
\usepackage{times}
\usepackage{xcolor}
\usepackage{longtable}
\usepackage[]{amsmath}
\usepackage[]{dsfont}
\usepackage[]{amsfonts}
\usepackage[]{amssymb}
\usepackage[]{stmaryrd}
\usepackage[]{mathrsfs}
\usepackage{hyperref}
\usepackage{url}
\usepackage{bm}
\usepackage[encapsulated]{CJK}

\title{Precessing jet nozzle connecting to a spinning black hole in M87
}

\author{Yuzhu Cui$^{1,2,3,4}$\thanks{yuzhu\_cui77@163.com},
Kazuhiro Hada$^{3,4}$,
Tomohisa Kawashima$^{5}$,
Motoki Kino$^{6,4}$,
Weikang Lin$^{7,2}$,
Yosuke Mizuno$^{2,8,9}$,
Hyunwook Ro$^{10,11}$,
Mareki Honma$^{4,12}$,
Kunwoo Yi$^{13}$,
Jintao Yu$^{14}$,
Jongho Park$^{10,15}$,
Wu Jiang$^{16,17}$,
Zhiqiang Shen$^{16,17}$,
Evgeniya Kravchenko$^{18,19}$,
Juan-Carlos Algaba$^{20}$,
Xiaopeng Cheng$^{10}$,
Ilje Cho$^{21,10}$,
Gabriele Giovannini$^{22,23}$,
Marcello Giroletti$^{23}$,
Taehyun Jung$^{10,24}$,
Ru-Sen Lu$^{16,17,25}$,
Kotaro Niinuma$^{26,27}$,
Junghwan Oh$^{28}$,
Ken Ohsuga$^{29}$,
Satoko Sawada-Satoh$^{30}$,
Bong Won Sohn$^{10,24,11}$,
Hiroyuki R. Takahashi$^{31}$,
Mieko Takamura$^{12,4}$,
Fumie Tazaki$^{32}$,
Sascha Trippe$^{13,33}$,
Kiyoaki Wajima$^{10,24}$,
Kazunori Akiyama$^{34,35,36}$,
Tao An$^{16}$,
Keiichi Asada$^{15}$,
Salvatore Buttaccio$^{23}$,
Do-Young Byun$^{10,24}$,
Lang Cui$^{37,17}$,
Yoshiaki Hagiwara$^{38}$,
Tomoya Hirota$^{3,4}$,
Jeffrey Hodgson$^{39}$,
Noriyuki Kawaguchi$^{4}$,
Jae-Young Kim$^{40,25}$,
Sang-Sung Lee$^{10,24}$,
Jee Won Lee$^{10}$,
Jeong Ae Lee$^{10}$,
Giuseppe Maccaferri$^{23}$,
Andrea Melis$^{41}$,
Alexey Melnikov$^{42}$,
Carlo Migoni$^{41}$,
Se-Jin Oh$^{10}$,
Koichiro Sugiyama$^{43}$,
Xuezheng Wang$^{16}$,
Yingkang Zhang$^{16}$,
Zhong Chen$^{16,17}$,
Ju-Yeon Hwang$^{10}$,
Dong-Kyu Jung$^{10}$,
Hyo-Ryoung Kim$^{10}$,
Jeong-Sook Kim$^{10,44}$,
Hideyuki Kobayashi$^{3}$,
Bin Li$^{16,17}$,
Guanghui Li$^{37}$,
Xiaofei Li$^{37}$,
Zhiyong Liu$^{37}$,
Qinghui Liu$^{16,17}$,
Xiang Liu$^{37}$,
Chung-Sik Oh$^{10}$,
Tomoaki Oyama$^{4}$,
Duk-Gyoo Roh$^{10}$,
Jinqing Wang$^{16,17}$,
Na Wang$^{37,17}$,
Shiqiang Wang$^{37}$,
Bo Xia$^{16}$,
Hao Yan$^{37}$,
Jae-Hwan Yeom$^{10}$,
Yoshinori Yonekura$^{45}$,
Jianping Yuan$^{37}$,
Hua Zhang$^{37}$,
Rongbing Zhao$^{16,17}$,
Weiye Zhong$^{16,17}$
}

\begin{document}

\maketitle

\begin{affiliations}
\item Research Center for Intelligent Computing Platforms, Zhejiang Laboratory, Hangzhou 311100, China
\item Tsung-Dao Lee Institute, Shanghai Jiao Tong University, 520 Shengrong Road, Shanghai 201210, China
\item Astronomical Science Program, The Graduate University for Advanced Studies, SOKENDAI, 2-21-1 Osawa, Mitaka, Tokyo 181-8588, Japan
\item Mizusawa VLBI Observatory, National Astronomical Observatory of Japan, 2-12 Hoshigaoka, Mizusawa, Oshu, Iwate 023-0861, Japan
\item Institute for Cosmic Ray Research, The University of Tokyo, 5-1-5 Kashiwanoha, Kashiwa, Chiba 277-8582, Japan
\item Kogakuin University of Technology \& Engineering, Academic Support Center, 2665-1 Nakano-machi, Hachioji, Tokyo 192-0015, Japan 
\item South-Western Institute For Astronomy Research, Yunnan University, Kunming 650500, China
\item School of Physics and Astronomy, Shanghai Jiao Tong University, Shanghai 200240, China 
\item Institut f\"ur Theoretische Physik, Goethe-Universit\"at Frankfurt, Max-von-Laue-Stra{\ss}e 1, D-60438 Frankfurt am Main, Germany
\item Korea Astronomy \& Space Science Institute, Daedeokdae-ro 776, Yuseong-gu, Daejeon 34055, Republic of Korea
\item Department of Astronomy, Yonsei University, Yonsei-ro 50, Seodaemun-gu, Seoul 03722, Republic of Korea
\item Department of Astronomy, Graduate School of Science, The University of Tokyo, 7-3-1, Hongo, Bunkyo, Tokyo 113-0033, Japan
\item Department of Physics and Astronomy, Seoul National University, Gwanak-gu, Seoul 08826, Republic of Korea
\item Department of Information Countermeasure, Air Force Early Warning Academy, Wuhan 430019, China
\item Institute of Astronomy and Astrophysics, Academia Sinica, Hilo, HI 96720, USA
\item Shanghai Astronomical Observatory, Chinese Academy of Sciences, 80 Nandan Road, Shanghai 200030, China
\item Key Laboratory of Radio Astronomy, Chinese Academy of Sciences, Beijing 100101, China
\item Moscow Institute of Physics and Technology, Dolgoprudny, Institutsky per., 9, Moscow 141700, Russia
\item Lebedev Physical Institute of the Russian Academy of Sciences, Leninsky prospekt 53, 119991 Moscow, Russia
\item Department of Physics, Faculty of Science, Universiti Malaya, 50603 Kuala Lumpur, Malaysia
\item Instituto de Astrof\'{\i}sica de Andaluc\'{\i}a - CSIC, Glorieta de la Astronom\'{\i}a s/n, E-18008 Granada, Spain
\item DIFA Bologna University via P. Gobetti 93/2, I-40129 Bologna, Italy
\item INAF-Istituto di Radioastronomia, Via P. Gobetti 101, I-40129 Bologna, Italy
\item University of Science and Technology, Gajeong-ro 217, Yuseong-gu, Daejeon 34113, Republic of Korea
\item Max-Planck-Institut f\"ur Radioastronomie, Auf dem H\"ugel 69, D-53121 Bonn, Germany
\item Graduate School of Sciences and Technology for Innovation, Yamaguchi University, 1677-1 Yoshida, Yamaguchi, Yamaguchi 753-8511, Japan
\item The Research Institute for Time Studies, Yamaguchi University, Yoshida 1677-1, Yamaguchi, Yamaguchi 753-8511, Japan
\item Joint Institute for VLBI ERIC, Oude Hoogeveensedijk 4, 7991 PD Dwingeloo, The Netherlands
\item Center for Computational Sciences, University of Tsukuba, Ten-nodai, 1-1-1 Tsukuba, Ibaraki 305-8577, Japan
\item Graduate School of Science, Osaka Metropolitan University, 1-1 Gakuen-cho, Naka-ku, Sakai, Osaka 599-8531, Japan
\item Department of Natural Sciences, Faculty of Arts and Sciences, Komazawa University, 1-23-1 Komazawa, Setagaya, Tokyo 154-8525, Japan
\item Tokyo Electron Technology Solutions Limited, 52 Matsunagane, Iwayado, Esashi, Oshu City, Iwate 023-1101, Japan
\item SNU Astronomy Research Center, Seoul National University, Gwanak-gu, Seoul 08826, Republic of Korea
\item National Radio Astronomy Observatory, 520 Edgemont Rd, Charlottesville, VA 22903, USA
\item Massachusetts Institute of Technology Haystack Observatory, 99 Millstone Road, Westford, MA 01886, USA
\item Black Hole Initiative at Harvard University, 20 Garden Street, Cambridge, MA 02138, USA
\item Xinjiang Astronomical Observatory, Chinese Academy of Sciences, Urumqi 830011, China
\item Toyo University, 5-28-20 Hakusan, Bunkyo-ku, Tokyo 112-8606, Japan
\item Department of Physics and Astronomy, Sejong University, Gwangjin-gu, Seoul 05006, Republic of Korea
\item Department of Astronomy and Atmospheric Sciences, Kyungpook National University, Daegu 702-701, Republic of Korea 
\item INAF - Osservatorio Astronomico di Cagliari, Via della Scienza 5, 09047, Selargius, CA, Italy
\item Institute of Applied Astronomy, Russian Academy of Sciences, Kutuzova Embankment 10, St. Petersburg, 191187, Russia
\item National Astronomical Research Institute of Thailand (Public Organization), 260 Moo 4, T. Donkaew, A. Maerim, Chiangmai, 50180, Thailand
\item National Astronomical Observatories, Chinese Academy of Sciences, 20A Datun Road, Chaoyang District, Beijing, China
\item Center for Astronomy, Ibaraki University, 2-1-1 Bunkyo, Mito, Ibaraki 310-8512, Japan
\end{affiliations}

\begin{abstract}
The nearby radio galaxy M87 offers a unique opportunity to explore the connections between the central supermassive black hole and relativistic jets. Previous studies of the inner region of M87 revealed a wide opening angle for the jet originating near the black hole~\cite{junor1999, hada2011n, walker2018, lu2023}. The Event Horizon Telescope resolved the central radio source and found an asymmetric ring structure consistent with expectations from General Relativity~\cite{eht12019}. With a baseline of 17 years of observations, there was a shift in the jet's transverse position, possibly arising from an eight to ten-year quasi-periodicity~\cite{walker2018}. However, the origin of this sideways shift remains unclear. Here we report an analysis of radio observations over 22 years that suggests a period of about 11 years in the position angle variation of the jet. We infer that we are seeing a spinning black hole that induces the Lense-Thirring precession of a misaligned accretion disk. Similar jet precession may commonly occur in other active galactic nuclei but has been challenging to detect owing to the small magnitude and long period of the variation.
\end{abstract}

\section*{Main}
\begin{figure}
\begin{center}
\includegraphics[width=0.45\textwidth]{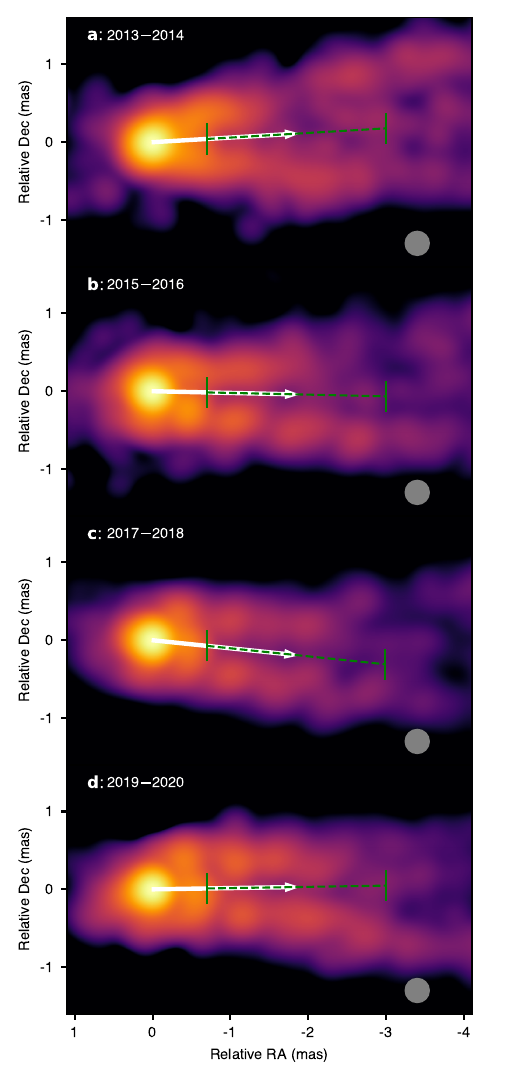}
\caption{\textbf{Structural evolution of M87 jet from 2013 to 2020.} \textbf{a-d}, The images are produced by stacking individual EAVN/VLBA Q-band images over every two years. The nearby years are indicated at the top-left corner: 2013--2014 (\textbf{a}); 2015--2016 (\textbf{b}); 2017--2018 (\textbf{c}); 2019--2020 (\textbf{d}). The grey-colored circle at the bottom-right corner of each panel indicates a common circular Gaussian beam with FHWM of 0.3\,mas. All images are rotated by $-18^\circ$. The white arrow in each panel indicates the jet PA averaged over a jet portion of 0.7--3.0\,mas from the core (indicated by the green dotted line) in the corresponding stacked images. For M87, BH mass $M_{\rm BH} = 6.5 \times 10^9\,M_{\odot}$~\cite{eht12019}, $1\,{\rm\,mas}\approx 250\,r_{\rm g} \approx0.08\,\rm pc$. Dec., declination; RA, right ascension.}
\label{fig:stru}
\end{center}
\end{figure}

To accurately trace the long-term morphological evolution of the M87 jet near the supermassive black hole (SMBH), we analyzed 170 Very Long Baseline Interferometry (VLBI) images of the M87 jet obtained with the East Asian VLBI Network (EAVN~\cite{cui2021}) and the Very Long Baseline Array (VLBA~\cite{walker2018}) at Q and K bands (referring to 43\,GHz and 22/24\,GHz, respectively) between 2000 and 2022 (\EXTFIG{fig:ssQ1} and \EXTFIG{fig:ssK1}). Part of the EAVN observations at K band was further connected to the telescopes in Italy and Russia (EATING~\cite{giovannini2023}). The detailed information of the data and the joined antenna are listed in \EXTTAB{tab:array} and \EXTTAB{tab:eating}. \FIG{fig:stru} presents a sequence of bi-yearly stacked EAVN/VLBA Q-band images obtained from 2013 to 2020. Besides the well-known persistent limb-brightened jet morphology~\cite{hada2016gmva}, one can see that the overall position angle (PA) of the jet direction near the core noticeably changes over the years. 

\begin{figure}
\begin{center}
\includegraphics[width=1\textwidth]{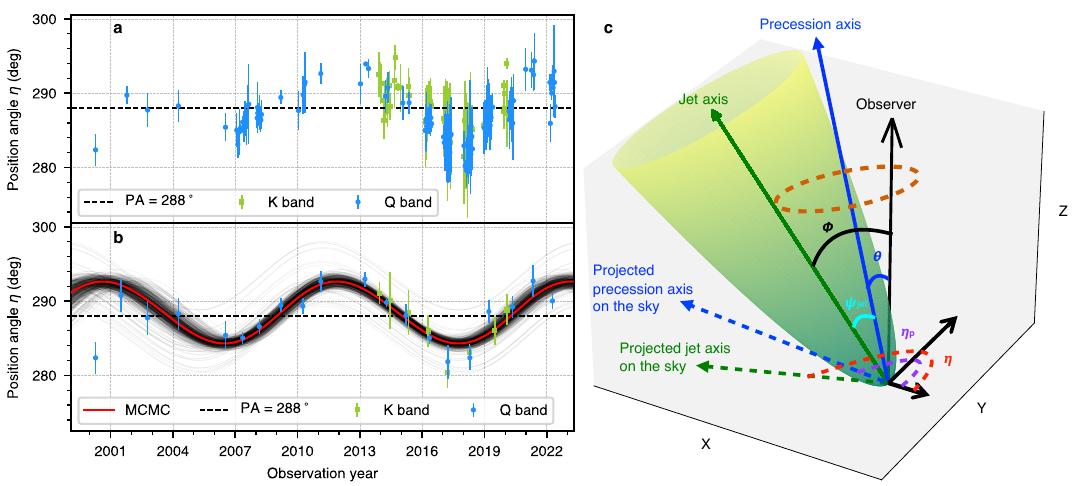}
\caption{\textbf{Time-dependence of the M87 jet PA from 2000 to 2022 and a schematic picture of the precession model.} \textbf{a, b}, Error bars represent the standard deviation derived from the Gaussian fitting. The green (blue) data points in \textbf{(a)} and \textbf{(b)} indicate the measured PA at K (Q) band. The horizontal dashed line represents the well-known jet PA in previous studies of M87~\cite{walker2018} in \textbf{(a)} and the best fit of $\eta_{\rm p}$ in \textbf{(b)}. For \textbf{(a)}, the measurements were conducted with 164 individual epochs within the jet distances 0.7--3.0\,mas (EAVN and VLBA Q band, EATING and VLBA K band) and 1.7--3.0\,mas (EAVN K band), respectively. For \textbf{(b)}, the results were obtained from the yearly binned images with all 170 epochs. Owing to the stacking procedure according to the observing frequency, the measured region for EATING and VLBA K band data is within 1.7--3.0\,mas core separation (Methods). The red line is derived from the best-fit precession model parameters (\TAB{tab:fitp}). The thin grey lines represent the statistical errors, which are randomly chosen from the MCMC samples. \textbf{(c)}, ({\it X,Y}) is the sky plane. As the jet precesses, the central axis (solid green arrow) of the parabolic jet (green surface) rotates with respect to the precession axis (solid blue arrow) along the trajectory indicated by the dotted orange circle. The surface traced out by the jet's central axis is dubbed the ``precession cone''. The dotted blue (green) arrow represents the projected precession (jet) axis. The PA of the projected jet axis $\eta$ (red dotted arc), the PA of the projected precession axis $\eta_{\rm p}$ (purple dotted arc), the half opening angle of jet precession cone $\psi_{\rm jet}$ (cyan arc), and the angle between the LOS ({\it Z} axis) and the jet (precession) axis $\phi$ ($\theta$) are labelled accordingly.
\label{fig:pa}}
\end{center}
\end{figure}

\FIG{fig:pa}-\textbf{(a)} displays the time evolution of the jet central PA averaged over distances of 0.7--3.0\,milli-arcseconds (mas) measured for 164 individual epochs after excluding 6 epochs with poor quality (Methods). Although the error bars of individual data points are relatively large, the ensemble of 164 measurements clearly reveals a systematic year-scale oscillation of the jet PA with a peak-to-peak amplitude of $\sim10^\circ$ centered at PA$\sim288^\circ$. Note that the parsec-scale jet of M87 is known to exhibit short-term (weekly and monthly) structural variations for various reasons, such as (1) episodic ejections of new jet components~\cite{hada2014}, (2) bulk flow acceleration along the jet~\cite{park2019v}, and (3) hydrodynamical instabilities that make the jet fluctuated transversely~\cite{ro2023}. Additionally, in combination with inhomogeneous image dynamic ranges among different epochs, these temporal effects cause large scatters in the measured PA among individual epochs within the same year. Hence, to smooth out short-term ($<$1-year scale) temporal fluctuations and then highlight the long-term global systematic evolution of the jet base, we produced a sequence of yearly-binned images by averaging multiple images over every single year. As shown in \FIG{fig:pa}-\textbf{(b)}, the yearly-binned evolution of the jet PA obtained from the stacked images displays clear quasi-sinusoidal variations as a function of observing year $t$. 

To characterize the periodic oscillation of the jet nozzle on the sky plane, here we introduce a simple model of the precessing solid-body cone in the three-dimensional space (\FIG{fig:pa}-\textbf{(c)}). The observed jet PA is identified as the angle of the jet axis projected on the sky, $\eta$, which is related to the intrinsic properties of the jet precession by applying a sequence of rotation matrices from the jet to the observer frame~\cite{caproni2004a} (Methods). We note that the angular velocity of precession is expected to be non-relativistic and the jet portion considered in the present analysis (0.7--3.0\,mas corresponding to the de-projected distances around 600--2,500\,$r_{\rm g}$ for a viewing angle $\theta = 17.2^\circ$~\cite{walker2018}, where $r_{\rm g}=GM_{\rm BH}/c^2$ is the gravitational radius, $M_{\rm BH}$ is the BH mass, $G$ is the gravitational constant, and $c$ is the speed of light) is in the weak gravity region. Therefore, the effect of relativistic time dilation is considered to be negligible in our present modeling. We perform a likelihood analysis using the Markov Chain Monte Carlo (MCMC) algorithm to a time series of the jet PA obtained from the yearly-binned jet images between 2006 to 2022. The best fitting result of $\eta$ is shown in \FIG{fig:pa}-\textbf{(b)} with a red thick line. The results of each parameter are listed in \TAB{tab:fitp}. 

The long-term PA data are well matched by the jet precession model with a best-fit reduced Chi-squared value $\hat{\chi}^2$ of $1.2$. Note that two cycles are still not definitive to conclude a periodicity considering the possible effect of red noise. Nevertheless, the good agreement between the observations and the jet precession model and the putative periodicity of the jet PA variation casts doubts on alternative scenarios like temporal oscillations by instabilities~\cite{mizuno2007, mizuno2012, walker2018} (see more in Methods). In our fiducial analysis, we include some prior of the viewing angle $\phi$ to break the degeneracy between the half opening angle $\psi_{\rm jet}$ of the jet precession cone and the angle $\theta$ between the precession axis and the line of sight (LOS), resulting in $\psi_{\rm jet}=(1.25\pm0.18)^\circ$. The deduced $|\omega_{\rm p}|=(0.56\pm0.02)$\,rad/year corresponds to a precession period of $T_{\rm prec}^{\rm jet} = (11.24\pm0.47)$\,years, which is comparable to the $8-10$ year quasi-periodicity reported in Ref.~\cite{walker2018}. The periodicity of the PA variation is robust regardless of whether we include the prior of $\phi$, whether we only use a subset of data (that at Q band), and whether we include the earlier data (from 2000 to 2004) with poor quality (Methods, \EXTFIG{fig:mcmc1}, \EXTTAB{tab:mcmc1}, \EXTTAB{tab:mcmc2}, \EXTTAB{tab:mcmc3}). As the jet precesses, the mean value of the jet viewing angle $\phi$ oscillates between $16^\circ$ and $18.5^\circ$ with an uncertainty of $\sim 2^\circ$ each year. The inferred evolution of $\phi$ is shown in \EXTFIG{fig:mcmc-phi} and its values for some selected years are listed in \EXTTAB{tab:mcmc-phi}.

The observed periodic PA variation in the M87 jet is likely triggered by certain physical and steady processes, and Lense-Thirring (LT) precession of a tilted accretion disk with respect to the SMBH spin~\cite{lense1918} is a promising origin. In fact, since the matter accreting onto the SMBH is insensitive to the BH spin direction, a certain misalignment between the angular momentum vector of the accretion disk and that of the SMBH spin is expected to commonly exist in active galactic nuclei (AGNs) with the level of misalignment depending on how exactly the SMBH became part of the system~\cite{fragile2007, mcKinney2013}. This configuration can generate LT precession of the accretion disk caused by the frame-dragging force of a spinning BH~\cite{lense1918}, which is expected to propagate to the jet through the tight coupling between the jet and the accretion disk~\cite{liska2018, mcKinney2013}.

\begin{figure}
\begin{center}
\includegraphics[width=.99\textwidth]{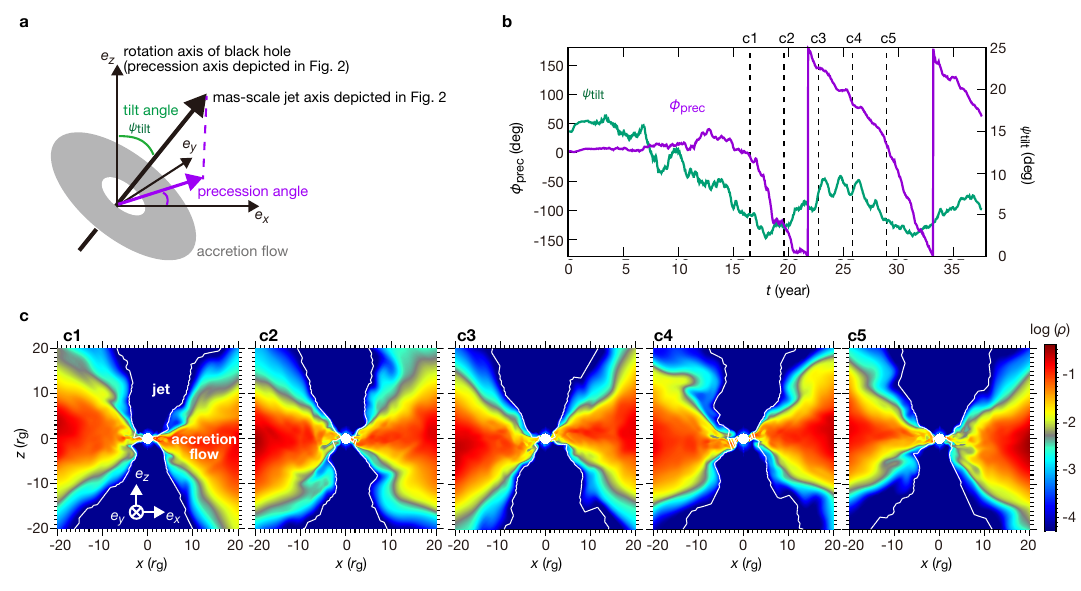}
\caption{\textbf{GRMHD simulation.} \textbf{(a)}, Configuration of the BH spin axis and mas-scale jet axis. $\psi_{\rm tilt}$ is the tilt angle of the mas-scale jet axis with respect to the BH spin axis ({\it z} axis). $\phi_{\rm prec}$ is the projected angle of the mas-scale jet axis relative to the {\it x} axis, which traces the precession of accretion flow. Here, the mas-scale jet axis is almost aligned with the rotation axis of the accretion flow and slow outflow due to the collimation effect while the jet axis is aligned with BH spin axis near the event horizon~\cite{mcKinney2013, liska2018}. \textbf{(b)}, Evolution of $\psi_{\rm tilt}$ and $\phi_{\rm prec}$ as a function of time evaluated at $r = 20\,r_{\rm g}$. Five time points, $t_1=16.23$\,years (c1), $t_2=19.27$\,years (c2), $t_3=22.32$\,years (c3), $t_4=25.36$\,years (c4) and $t_5=28.41$\,years (c5), are indicated with the dashed lines. \textbf{(c)}, Snapshots of mass density $\rho$ of the GRMHD simulation results corresponding to the five-time points indicated in the subplot \textbf{(b)}, where $\rho$ is measured in the fluid-rest frame. We set $M_{\rm BH}= 6.5 \times 10^9\,M_{\odot}$ for the conversion of the time unit in GRMHD simulation $r_{\rm g}/c$ to year. The total simulated time is $3.7 \times 10^4\,r_{\rm g}/c$, namely 37.5\,years. We take a spin parameter $a_*=0.9375$. The white contours depict the surface of magnetization $\sigma (\equiv B^2/\rho c^2) = 1$, where $B$ is the magnetic field strength in the fluid-rest frame. We define the regions $\sigma > 1$ (that is, the region near the $z$ axis inside the contour) as the relativistic jets. 
\label{fig:grmhd}}
\end{center}
\end{figure}

Extensive general relativistic magnetohydrodynamics (GRMHD) simulations exploring misaligned systems have demonstrated that the majority of the accretion disk coherently experiences LT precession~\cite{fragile2007, liska2018, white2019, chatterjee2020, ressler2023} and the jet indeed precesses in phase with the disk~\cite{mcKinney2013, liska2018}. Adopting the SMBH mass of M87, the pioneering work by Refs.~\cite{fragile2007, white2019} reproduced a precession period at the same order of magnitude as that deduced in this work. Here we further develop and conduct our GRMHD simulations with settings closely resembling the M87 system and successfully recover the disk/jet precession in a tilted disk-BH system over almost two cycles by using \texttt{UWABAMI} code~\cite{takahashi2016, kawashima2023} (\FIG{fig:grmhd} and Methods). A steady precession with a period consistent with the observations is revealed after the simulation converges at around $t=16$\,years, as indicated in the panel {\bf(b)} in \FIG{fig:grmhd}. While we assume a spin parameter $a_*=0.9375$, the exact relation between the precession period and the BH spin is sensitive to the morphology of the disk~\cite{liu_2002, fragile2007}. 

The presence of LT precession indicates that the M87 central SMBH is spinning, which is essential to produce an energetic jet via the Blandford-Znajek mechanism~\cite{blandford1977}. However, the magnitude of spin is sensitive to the size of the disk according to $T_{\rm prec}={\pi c^3 r_{\textsc{lt}}^3}/{(G^2a_*M_{\rm BH}^2)}$ in the weak-field limit \cite{wilkins1972, caproni2007, fragile2007}. On the other hand, the effective radius is $r_{\textsc{lt}}\sim15\,r_{\rm g}$ for a maximally spinning SMBH, which suggests a compact disk region that undergoes coherent LT precession and motivates us to adopt a small disk in simulations. The small value of $\psi_{\rm jet}$ implies a small misalignment between the M87 jet at mas scales and the SMBH spin. Due to the tight relation between the jet and accretion disk~\cite{mcKinney2013,liska2018}, it further suggests a slight tilt of the accretion disk with respect to the SMBH spin. Such a configuration may naturally arise if the M87 central SMBH grew mainly through accretion~\cite{scheuer1996, natarajan1999}. However, the disk could have a finite tilt angle as the disk orientation can vary with radius~\cite{liska2018,white2019}. In that case, since the bright side of the ring-like structure detected by Event Horizon Telescope (EHT) is connected to the relativistic jet base, the PA change of it~\cite{wielgus2020} may synchronize with the PA change seen in the mas-scale jet. The recent Global Millimeter VLBI Array observations at 86\,GHz, which successfully detected both the ring-like structure and jet, well fill the spatial gaps between mas and micro-arcsecond scales~\cite{lu2023}. Further accumulating multi-year, multi-wavelength VLBI images is crucial to seamlessly connect the dynamic evolution of the structure from the emission surrounding the BH to the launching jet~\cite{chatterjee2020}.


\begin{table*}[htbp]
 \begin{tabular*}{1\textwidth}{@{\extracolsep{\fill}}l|llll}
  \hline
    \hline
     Parameters             & Definition                                                              &     Value$^{a}$                       &  Unit   \\
    \hline
    \hline
    $t_0$                   & Reference time                                                          &  $2014.82\pm0.15$             &    year   \\
    $\eta_{\rm p}$          & PA of the precession axis                                          &  $ 288.47\pm0.27$                &    deg  \\
    $|\omega_{\rm p}|$      & Angular velocity of precession                                          &  $0.56\pm0.02$                  &   rad/year\\
    $\psi_{\rm jet}$        & Half opening angle of the precession cone                               &  $1.25\pm0.18$                  &    deg  \\
    $\theta$                & Angle between the precession axis and LOS                &  $17.21\pm1.74$                &    deg  \\
    \hline
\end{tabular*}
\caption{\textbf{Summary of the parameters in the precession model.} The configuration of these free parameters refers to \FIG{fig:pa}-\textbf{(c)}. $^{a}$: The values correspond to the means of the MCMC samples with standard deviations. The period of precession, $T_{\rm prec}^{\rm jet}=2\pi/\omega_{\rm p} = (11.24\pm0.47)$\,years. The sense of precession (the sign of $\omega_{\rm p}$) can not be determined by the observed PA of the jet axis $\eta$ in this work. 
}\label{tab:fitp}
\end{table*}

\pagebreak
\begin{methods}
In what follows we describe the methods and assumptions employed to derive our results in Main.
\subsection{Summary of observations and data}
\label{sec:data}
The data presented in this work are taken from EAVN and VLBA at 22/24/43\,GHz and EATING (a joint array of EAVN and radio telescopes in Italy and Russia) at 22\,GHz. The primary data among them are the ones obtained with EAVN and VLBA at 43\,GHz since these data provide finer angular resolution, imaging sensitivity, and observing cadences to monitor the jet base, while the other data further complement the 43\,GHz data (i.e., higher resolution with EATING; larger field-of-view at 22\,GHz). In total, the VLBI data we used for imaging analysis include 119 epochs from EAVN, 4 epochs from EATING, and 47 epochs from VLBA. Basic information on the data from each of these VLBI arrays is summarized in \EXTTAB{tab:array}. Divided by observing frequency, there are 56 epochs at 22/24\,GHz and 114 epochs at 43\,GHz. The total observation period covered by these observations is from 2000 April to 2022 May. All the individual data are summarized in Supplementary Information.

\subsection{Notes on EAVN and EATING data} Since 2013 we have been regularly monitoring the pc-scale jet of M87 with EAVN, a joint VLBI network in East Asia. In each year the EAVN monitoring observations were performed mainly from December to June with sampling intervals ranging from a few days to a month. While the EAVN observations until 2016 were conducted with a joint array of KVN (Korean VLBI Network, Korea) and VERA (VLBI Exploration of Radio Astrometry, Japan), namely KaVA, from 2017 more stations in East Asia joined the network, enhancing the overall array performance. The angular resolution of only KaVA is 1.26\,mas at 22\,GHz and 0.63\,mas at 43\,GHz. The default array configurations from 2017 were KaVA+Tianma+Nanshan at 22\,GHz and KaVA+Tianma at 43\,GHz, respectively. This achieves a maximum angular resolution of 0.55\,mas at 22\,GHz and 0.63\,mas at 43\,GHz. Additionally, part of our EAVN 22\,GHz observations were further connected to the telescopes in Italy (Medicina and Sardinia stations) or Russia (Badary station). We call this global network an ``EATING VLBI array" and extend our maximum baseline lengths from 5,078\,km to $\sim$10,000\,km, resolving the regions closer to the BH at a resolution down to 0.27\,mas (mainly in the east-west direction). We performed 4 epochs of EATING VLBI sessions between 2017 and 2020 (see \EXTTAB{tab:eating} for details). 

Each of the EAVN/EATING sessions was made in a 5--7-hour continuous run at a data recording rate of 1\,Gbps (a total bandwidth of 256\,MHz). Only left-hand circular polarization was recorded. All the data were correlated at the Daejeon hardware correlator installed at Korea Astronomy and Space Science Institute (KASI). The correlated data were calibrated in the standard manner of VLBI data reduction procedures and under the guideline of EAVN data reduction~\footnote{\url{https://radio.kasi.re.kr/eavn/data_reduction.php}}. The initial calibration of visibility amplitude, phase, and bandpass was performed with the {\tt AIPS} software package~\cite{greisen2003}. The subsequent imaging~\cite{hogbom1974} and self-calibration were performed with the {\tt Difmap} software~\cite{shepherd1994}.

\subsection{Notes on VLBA archival data} To expand the time coverage of our study, we additionally reanalyzed VLBA archival data obtained between 2000 and 2020. The VLBA data between 2006 and 2018 were part of a dedicated M87 monitoring program~\cite{walker2018}, while the data before 2006 were sparsely sampled with relatively lower imaging quality. There are 3 sessions observed at 24\,GHz and 44 sessions observed at 43\,GHz. The recording rate ranges from 128\,Mbps to 2,048\,Mbps depending on the sessions. Both left and right circular polarizations were recorded for most of these sessions. More detailed information for individual epochs can be found in Refs.~\cite{walker2018, ly2004, ly2007}. The data reduction process follows the standard process of VLBA data reduction. After the phase and amplitude calibration in the {\tt AIPS} software package, we did self-calibration and final imaging in the {\tt Difmap} software. The angular resolution of the VLBA image is around 0.40\,mas at 24\,GHz and 0.23\,mas at 43\,GHz, respectively. 

\subsection{Measurement of jet position angle}
We quantified the jet PA of M87 near the core in the following procedures. First, to reduce the effects from the shape of beam sizes, all images for the individual epochs are restored with a circular beam with sizes of 0.3\,mas for VLBA-43GHz and EATING-22\,GHz, 0.5\,mas for EAVN-43GHz and VLBA-24GHz data, and 1.2\,mas for EAVN-22GHz data, respectively. For each image, we then made circular slices (centered on the core) of the jet every 0.1\,mas from $r_{\rm start}$ (the staring distance of slicing) to 3.0\,mas along the jet, and integrated them over all sliced distances. Here $r_{\rm start}$ was set to at least 1.4 times the beam size of each image (0.7\,mas for VLBA-43GHz/EATING-22GHz/VLBA-24GHz/EAVN-43GHz images, while 1.7\,mas for EAVN-22GHz images) so that we can avoid the influences from the bright core~\cite{mertens2016}. The integration over a certain distance improves the significance and reduces the weight of the temporal emission caused by hydrodynamical instabilities in individual epochs. Then the integrated slice for each epoch was fitted with two or three Gaussian components owing to the well-known double-ridge or triple-ridge mas-scale jet profile of M87~\cite{hada2016gmva, hada2017galax}. Finally, we defined the middle of the outer two Gaussian peaks as the central PA of the M87 jet for each epoch. The errors come from two parts: 1) image noise which can be ignored due to the high enough significance after integration within the innermost region, and 2) Gaussian fitting errors output from the program. As a result, we adopted the Gaussian fitting errors as the error bars in Fig.~2-\textbf{(a)(b)}. Note that the core shift between 22/24\,GHz and 43\,GHz is $\sim$0.03\,mas according to Ref.~\cite{hada2011n}, which is negligible when we determine the PA in Fig.~2-\textbf{(a)(b)}. Through these procedures, we obtained PA (and its uncertainty) for the innermost region 0.7--3.0\,mas in 164 individual epochs as shown in Fig.~2-\textbf{(a).} Note that there are 6 epochs excluded in this individual epoch analysis due to the poor data quality as marked with $^\star$ in Supplementary Information, which may lead to some apparent differences between individual and stacked analysis, like the data point in 2001 shown in Fig.~2-\textbf{(a)(b)}.

In addition to the above-mentioned analysis on individual epochs, we also conducted a similar procedure for yearly stacked images with all 170 images. Before stacking, all the images are restored with a common circular Gaussian beam according to observing frequencies: 0.5\,mas for Q-band and 1.2\,mas for K-band data. Namely, the actual distance is 0.7--3.0\,mas covered by Q-band data while 1.7--3.0\,mas by K-band data. The stacked images have relatively higher signal-to-noise ratios compared with the individual epochs and smooth out the short-term variation which is better to trace the yearly variation seen in the M87 jet. Indeed, the analysis from the stacked images reveals the year-scale quasi-sinusoidal evolution of jet PA more clearly, although the trend before 2005 is less definitive due to the lack of multiple images within each year, in which case a single PA measurement may suffer from short-term temporal fluctuations as mentioned above. It should also be noted that the evolution of the measured PA at Q and K bands are in good agreement with each other, indicating the achromatic nature of the observed long-term jet base oscillation. The full sequences of the yearly stacked structure at Q and K bands are shown in \EXTFIG{fig:ssQ1} and \EXTFIG{fig:ssK1}, respectively. Note that in several years, including 2000, 2001, 2002, 2006, 2009, 2011, 2014, and 2015, there are only one or two epochs in those years. Hence the apparent jet structure is much more knotty than that in other years, the results of which may suffer more uncertainties from the short-term structural variations and data quality compared with data in other observing years.

\subsection{Precession model}
\label{sec:model}
In this section, we describe the precession model and the process of obtaining the relation between observed data and intrinsic physical parameters. A sequence of rotation matrices {\it\textbf{$\textbf{R}_{\rm i}(\xi)$}} are applied to obtain a vector represented in the jet frame to that in the observer frame, where $i$ denotes $x$, $y$ or $z$, and $\xi$ indicates the counterclockwise rotated angle with respect to the axis $i$. Observations show that on average the jet PA is $\sim288^\circ$ with a jet (projected) viewing angle of about $17^\circ$~\cite{walker2018}. The PA variation alone cannot determine the direction of the jet precession. For simplicity, we first assume a clockwise precession with respect to the precession axis. A counter-clockwise case equally fits the data well with a suitable shift of the reference time $t_0$.

In the jet frame, the unit vector of the jet symmetric axis in a Cartesian coordinate system can be expressed as $\vec{\textbf{j}}=[0,0,1]$. In the precession frame where the precession axis is along the ${z}$ axis, assuming a precession angular velocity of $\omega_{\rm p}$, the jet precesses with respect to the ${z}$ axis with an angle of $\omega_{\rm p}(t-t_0)$ in a time difference of $(t-t_0)$. The half-opening angle of the jet precession cone is $\psi_{\rm jet}$. Then, the symmetric jet axis represented in the precession frame is obtained by the following operations to $\vec{\textbf{j}}$, 
\begin{equation}\label{jp}
\vec{\bm{j}}_{\rm p}= \bm{R}_{\rm z}(-\omega_{\rm p}(t-t_0))\bm{R}_{\rm y}(\psi_{\rm jet}) \vec{\bm{j}}\,.
\end{equation} 

In the observer frame, the angle between the precession axis and the ${z}$ axis (the LOS) is $\theta$, corresponding to a rotation of $\bm{R}_{\rm y}(\theta)$. The projection of the precession axis in the $x$-$y$ plane makes an angle of $\eta_{\rm p}$ with the ${x}$ axis, corresponding to another rotation of $\bm{R}_{\rm z}(\eta_{\rm p})$. Therefore, the jet symmetric axis presented in the observer frame can be obtained by applying $\bm{R}_{\rm y}(\theta)$ and $\bm{R}_{\rm z}(\eta_{\rm p})$ successively to that represented in the precession frame, namely,
\begin{equation}\label{jo}
\vec{\bm{j}}_{\rm o}= \bm{R}_{\rm z}(\eta_{\rm p})\bm{R}_{\rm y}(\theta)\vec{\bm{j}}_{\rm p}\,.
\end{equation} 

By combing Equation~\eqref{jp} and Equation~\eqref{jo}, the jet axis components in the observer frame ($j_{\rm xo}(t),\,j_{\rm yo}(t),\,j_{\rm zo}(t)$) can be written as:
\begin{equation}\label{jox}
j_{\rm xo}(t) = A\cos\eta_{\rm p}-B\sin\eta_{\rm p}\,,
\end{equation} 
\begin{equation}\label{joy}
j_{\rm yo}(t) = A\sin\eta_{\rm p}+B\cos\eta_{\rm p}\,,
\end{equation} 
\begin{equation}\label{joz}
j_{\rm zo}(t) = -\sin(\theta)\cos(-\omega_{\rm p}(t-t_0))\sin(\psi)+\cos(\theta)\cos(\psi)\,,
\end{equation} 
where,
\begin{equation}\label{A}
A = \cos(\theta)\cos(-\omega_{\rm p}(t-t_0))\sin(\psi)+\sin(\theta)\cos(\psi)\,,
\end{equation} 
\begin{equation}\label{B}
B = \sin(-\omega_{\rm p}(t-t_0))\sin(\psi)\,,
\end{equation} 
The observed PA $\eta(t)$ is the projected jet axis in the observer frame at observing time point $t$, which can be expressed as:
\begin{equation}\label{pa}
\eta(t)= \arctan{(\frac{j_{\rm yo}(t)}{j_{\rm xo}(t)})}\,. 
\end{equation}
The jet viewing angle at time point $t$ is $\phi(t)$ which can be given by:
\begin{equation}\label{va}
\phi(t) = \arcsin{\sqrt{j_{\rm xo}(t)^2+j_{\rm yo}(t)^2}}\,. 
\end{equation}

There are five free parameters: $t_0$, $\eta_{\rm p}$, $\omega_{\rm p}$, $\psi_{\rm jet}$ and $\theta$. 
The resultant predicted curve of $\eta(t)$ exhibits a quasi-sinusoidal variation with time, with the deviation from a sinusoidal curve depending on the relation between $\psi_{\rm jet}$ and $\theta$. Namely, there are four situations: (1) When $\theta=90^\circ$ and $\psi_{\rm jet}<90^\circ$, the $\eta(t)$ curve is exactly sinusoidal; (2) When $\psi_{\rm jet}<\theta<90^\circ$ but $\psi_{\rm jet}$ is not close to $\theta$, the $\eta(t)$ curve only slightly deviates from a sinusoid with a small skewness; (3) When $\psi_{\rm jet}$ is close to but still smaller than $\theta$, a large skewness appears. The direction of the skewness depends on the sense of $\omega_{\rm p}$; (4) When $\psi_{\rm jet}>\theta$, the jet rotates around LOS as viewed in the 2D projected plane and $\eta(t)$ continuously increases or decreases depending on the sense of $\omega_{\rm p}$. For the latter two cases the PA observation would deviate significantly from a sinusoidal curve, one can then determine all the five parameters in the precession model including the sense of $\omega_{\rm p}$ with the PA observation alone. Our case is however that with a very small skewness. Since only four parameters are needed to specify a sinusoidal curve and the peak-to-peak amplitude of $\eta(t)$ is determined by both $\psi_{\rm jet}$ and $\theta$, there is a degeneracy between $\psi_{\rm jet}$ and $\theta$ when only the PA observation is involved. The sense of $\omega_{\rm p}$ cannot be determined, either. In order to break the degeneracy between $\psi_{\rm jet}$ and $\theta$, we include in the fiducial analysis additional constraints of the jet viewing angle which we shall discuss in the following section.

Note that, this precession model is based on the solid body assumption, while this assumption eventually loses its validity at larger scales. Moreover, the mass density of M87 jet is low~\cite{ferrari1998, kino2004} and the SMBH in M87 is accreting at sub-Eddington rates~\cite{narayan1995, chatterjee2022}. In comparison to the sources that have high mass density jets and super-Eddington accretion (i.e., SS 433~\cite{hjellming1981}), the M87 jet at large scales is significantly more susceptible to the impacts of ambient environment~\cite{mcKinney2006}.

\subsection{A Bayesian analysis}
We describe here the likelihood function and prior that are used in the Bayesian analysis. According to the Bayes theorem, the posterior reads
\begin{equation}\label{eq:posterior}
    P(\bm{\lambda}|\bm{q})=\frac{\mathcal{L}(\bm{q}|\bm{\lambda})P(\bm{\lambda})}{P(\bm{q})}\,,
\end{equation}
where $\bm{\lambda}=(t_0,\eta_{\rm p},\omega_{\rm p},\psi_{\rm jet},\theta)$, $\bm{q}$ stands for observations, $\mathcal{L}(\bm{q}|\bm{\lambda})$ is the (joint) likelihood, $P(\bm{\lambda})$ is the prior, and $P(\bm{q})$ is the evidence which only serves as a constant normalization factor. We adopt uniform priors for $t_0$, $\eta_{\rm p}$, $\omega_{\rm p}$, $\psi_{\rm jet}$ and $\theta$ as displayed in \EXTTAB{tab:mcmc1}. 

The PA likelihood is assumed to be Gaussian and reads,
\begin{equation}\label{eq:likelihood_PA}
    -\ln\mathcal{L}_{\rm PA}=\sum_{i}\frac{(\eta_{\rm p}(t_{\rm i})-\eta_{\rm ob}(t_{\rm i}))^2}{2\sigma_{\rm i}^2}\,,
\end{equation}
where $\sigma_{\rm i}$ is the uncertainty of the observed PA at $t_{\rm i}$. Due to the poor quality of the VLBA data before 2006, the PAs are the observations from 2006 to 2022 at both 22/24 and 43\,GHz.

In addition to the PA observations, we consider the constraints on the jet viewing angle given in the literature. Since the jet is precessing, its viewing angle also varies with time. Therefore, when applying those constraints, we pay attention to the times when the observations are. We note that, in the precession model, the angle between LOS and precession axis $\theta$ is a different physical parameter from that between LOS and jet central axis $\phi$. Previous studies provided the constraints on $\phi$. Ref.~\cite{mertens2016} reported a viewing angle of $(17.2\pm3.3)^\circ$ based on the kinematic analysis with VLBA data observed in 2007, which gives the following Gaussian likelihood (up to a normalization constant)
\begin{equation}\label{eq:likelihood_VA1}
    -\ln\mathcal{L}_{\rm VA2007A}=\frac{(\phi_{2007.36}-17.2)^2}{2\times3.3^2}\,,
\end{equation}
where $\phi_{2007.36}$ is the jet viewing angle at $t=2007.36$.
Furthermore, the brightness ratio of the forward-jet to the counter-jet measured at the distance between 0.4 and 0.8\,mas from the core from VLBA data gives $(13\leq\phi_{2007.36}\leq27)^\circ$~\cite{mertens2016}. We represent this\ constraint by a piece-wise likelihood,
\begin{equation}\label{eq:likelihood_VA2}
    -\ln\mathcal{L}_{\rm VA2007B}=\left\{
    \begin{array}{ll}
         0\,, & 13^\circ\le\phi_{2007.36}\le27^\circ\,, \\
         \infty\,, & {\rm others}\,.
    \end{array}\right.
\end{equation}
By monitoring the fastest component (6\,c) with the Hubble Space Telescope (HST) from 1994.59 to 1998.55, Ref.~\cite{biretta1999} provided an upper limit $\phi_{1996.57}\le 19^\circ$. We represent this\ constraint by another piece-wise likelihood,
\begin{equation}\label{eq:likelihood_VA3}
    -\ln\mathcal{L}_{\rm VA1996}=\left\{
    \begin{array}{ll}
         0\,, &~\phi_{1996.57}\le19^\circ\,, \\
         \infty\,, & {\rm others}\,.
    \end{array}\right.
\end{equation}

We consider three cases to use the above-mentioned constraints on the jet viewing angle: 1) in Case I, we don't consider additional constraints on the jet viewing angle and only use PA data to perform the analysis; 2) in Case II, we use uniform distribution [0, 90]$^\circ$ for $\theta$ and put constraints on $\phi_{2007.36}=(17.2\pm3.3)^\circ$; 3) in Case III, in addition to Case II, we further consider the constraint on $13^\circ\le\phi_{2007.36}\le27^\circ$ and that on $\phi_{1996.57}\le 19^\circ$. More explicitly,
\begin{equation*}
    \begin{array}{ll}
         {\rm Case~I:} & \ln\mathcal{L}(\bm{q}|\bm{\lambda})=\ln\mathcal{L}_{\rm PA}\,, \\
         {\rm Case~II:} & \ln\mathcal{L}(\bm{q}|\bm{\lambda})=\ln\mathcal{L}_{\rm PA}+\ln\mathcal{L}_{\rm VA2007A}\,,\\
         {\rm Case~III:} & \ln\mathcal{L}(\bm{q}|\bm{\lambda})=\ln\mathcal{L}_{\rm PA}+\ln\mathcal{L}_{\rm VA2007A}+\ln\mathcal{L}_{\rm VA2007B}+\ln\mathcal{L}_{\rm VA1996}\,.
    \end{array}
\end{equation*}
We use the Python package {\tt EMCEE}~\cite{foreman2013} to explore the five free parameters with an MCMC sampler. We set the walker number to 32 and the iteration number to 10,000. 

For these three cases, the marginalized distributions of the parameters are presented in \EXTFIG{fig:mcmc1}-\textbf{(a)}. The results of $t_0$, $\eta_{\rm p}$, and $\omega_{\rm p}$ are insensitive to the constraints on the jet viewing angle. On the other hand, if only the PA data are used, there is a degeneracy between $\psi_{\rm jet}$ and $\theta$ as shown in Case I in \EXTFIG{fig:mcmc1}. This degeneracy is due to a geometrical effect, and for the same PA variation, the required precession half-opening angle is smaller when the precession axis is more aligned with LOS (smaller $\theta$). Such a degeneracy is broken when we apply some prior constraints on the jet viewing angle, which can be seen when we compared Case I to Case II or to Case III. It is worth pointing out that the conclusion of a small jet precession half-opening angle is insensitive to the constraints of $\phi$ adopted. Indeed, from the amplitude of the PA variation alone, we can already infer that the maximum half opening angle is $\psi_{\rm jet}^{\rm max}\sim5^\circ$ corresponding the case when $\theta=90^\circ$. Applying the constraints on the jet viewing angle at some specific years further constrains $\theta$ and reduces the value of $\psi_{\rm jet}$ to $\sim1^\circ$. To break the $\psi_{\rm{jet}}$-$\theta$ degeneracy, we adopt the results obtained from Case III as the final fitting results shown in the main text.

To check the robustness of our final results, we also performed analyses with four different data sets of PA as shown in \EXTFIG{fig:mcmc1}-\textbf{(b)}. In addition to the data observed from 2006 to 2022 at both Q and K bands applied in Case I -- Case III, we consider cases IV, V, and VI, whose likelihoods are the same as Case III except for the different data sets used in $\ln\mathcal{L}_{\rm PA}$. More explicitly, we have in Case IV the extended data observed from 2000 to 2022 at both Q and K bands, in Case V the data observed from 2006 to 2022 but only at Q band, and in Case VI the extended data observed from 2000 to 2022 but only at only Q band. As shown from the comparison of the parameter constraints through Case III to Case VI, our results are robust against whether the poor-quantity data before 2006 are added to the analysis or whether only the data at Q band are being used. The specifications of all cases are listed in \EXTTAB{tab:mcmc2} and all the MCMC fitting results are shown in \EXTTAB{tab:mcmc3}.

To access the goodness-of-fit of our precession model, we define the reduced Chi-squared value $\hat{\chi}^2$ as
\begin{equation}\label{eq:reduced-chi2}
    \hat{\chi}^2=\frac{\ln\mathcal{L}(\bm{q}|\bm{\lambda}_{\rm{best}})}{N_{\rm data}-N_{\rm param}}\,,
\end{equation}
where $\bm{\lambda}_{\rm best}$ is the best-fit model parameter vector, $N_{\rm data}$ is the number of data, and $N_{\rm param}$ is the number of model parameters. As shown in \EXTTAB{tab:mcmc3}, the $\hat{\chi}^2$ values are close to unity for most cases, indicating that our precession model well fits the observations. Exceptions are Cases IV and VI, where the data before 2006 are included and the $\hat{\chi}^2$ values are larger. However, the increase of the $\hat{\chi}^2$ for these two cases is only caused by one data point at $t=2000$. If that single data point is excluded from the analysis, the $\hat{\chi}^2$ drops back to around unity indicating the data point at $t=2000$ is an outlier. Compared with Ref.~\cite{ly2007}, the derived structure is consistent with previous work. However, the exact reason for the more southern PA with respect to the predicted trend is not very clear at this moment.

\subsection{GRMHD simulation}
\label{sec:grmhd}
We have carried out three-dimensional ideal GRMHD simulations of tilted accretion flows and relativistic jets around a spinning BH by using a GR-radiation-MHD code \texttt{UWABAMI}~\cite{takahashi2016}. For simplicity, we ignored the effect of the radiation (i.e., radiative force, cooling, and so on) for the simulation as is the case of the aligned disk simulation~\cite{kawashima2023}. We fixed the specific heat index $\gamma_{\rm heat} = 13/9$ because of the combination of assumption of non-relativistic protons $\gamma_{\rm heat} = 5/3$ and relativistic electrons $\gamma_{\rm heat} = 4/3$. The simulation is carried out up to $3.7\times 10^4 r_{\rm g}/c$.

The metric can be assumed to be fixed since the accretion rate is too low to affect the spacetime geometry. The GRMHD equations in the modified Kerr-Schild coordinate $(r_{\rm sim}, \theta_{\rm sim}, \phi_{\rm sim})$ are integrated. The magnitude of the dimensionless BH spin is set to be $a_* = 0.9375$~\cite{gammie2004}, which is close to the value of the maximum spin of a Kerr BH ($|a_* = 1|$).

We set the initial equilibrium torus with the tilt angle, which is the angle between the BH spin vector and the angular momentum vector of the torus, to be $\theta_{\rm sim} = \psi_{\rm tilt} = 15^\circ$. 
The direction of the BH spin vector is aligned with the direction of $\theta_{\rm sim} = 0^\circ$.
The inner edge $r_{\rm in}$ and the pressure (i.e., density) maximum of the initial torus $r_{\rm max}$ is set at $r_{\rm sim} =20\,r_{\rm g}$ and $33\,r_{\rm g}$, respectively. We note that larger $r_{\rm in}/r_{\rm max}$, which is a sensitive parameter governing the torus size, results in the smaller radius of the outer edge of the initial torus because the weaker pressure gradient force inside the torus is required for the dynamical equilibrium.
Here, $r_{\rm in}/r_{\rm max} \sim 0.6$ in our simulation is larger than a previous work $r_{\rm in}/r_{\rm max} = 0.5$~\cite{liska2018}.
As a consequence, the size of the accretion disk $r_{\rm disk}$, which is average in the disk mid-plane weighted by rest mass density~\cite{liska2018},
is initially $r_{\rm disk} \simeq 47\,r_{\rm g}$, i.e., a compact initial torus appears in our setup. A single poloidal magnetic flux loop with a vector potential $A_{\phi_{\rm sim}} \propto \max(\rho/\rho_{\rm max} - 0.2, 0)$ is embedded in the initial torus. Because (i) the initial torus is relatively compact, (ii) and located at moderately far radius from the BH, (iii) and the initial magnetic field is not so strong in the outer part of the initial torus, the resultant magnetic flux averaged in time during the precession phase (1.5--3.7)\,$\times10^4r_{\rm g}/c$ at the event horizon is $\phi_{\rm BH} \equiv \Phi_{\rm BH}/\sqrt{\dot M_{\rm BH}r_{\rm g}^2 c} \sim 17$, where mass accretion rate ${\dot M}_{\rm BH} \equiv \int^\pi_0 d{\theta_{\rm sim}} \int^{2\pi}_0 d\phi_{\rm sim} \sqrt{-g} \rho(r_{\rm sim}=r_{\rm g},\theta_{\rm sim}, \phi_{\rm sim}) u^{r_{\rm sim}} (r_{\rm sim} = r_{\rm g}, \theta_{\rm sim}, \phi_{\rm sim})$, $\Phi_{\rm BH} \equiv (1/2) \int^\pi_0 d{\theta_{\rm sim}} \int^{2\pi}_0 d\phi_{\rm sim} \sqrt{-g} B^{r_{\rm sim}} (r_{\rm sim} = r_{\rm g}, \theta_{\rm sim}, \phi_{\rm sim})$.
The magnetic field is evaluated and defined in the cgs-Gauss unit. 
This magnitude of magnetic flux is between the weakly magnetized disk state so-called SANE (Standard And Normal Evolution, $\phi_{\rm BH} \sim $ few--10~\cite{narayan2012, porth2019}) and the strongly magnetized disk state so-called MAD (Magnetically Arrested Disk, $20 \lesssim \phi_{\rm BH} \lesssim 60$~\cite{tchek2011, narayan2022}), and therefore, this intermediate state we adopted is sometimes called semi-MAD~\cite{white2020, anantua2020}.

The inner and outer boundaries of the simulation domain are set to be $r_{\rm in} = 1.18\,r_{\rm g}$ and $r_{\rm out} = 10^3\,r_{\rm g}$.
The simulation domain is divided into $(N_{r_{\rm sim}}, N_{\theta_{\rm sim}}, N_{\phi_{\rm sim}} ) = (200, 144, 96)$ meshes in $r_{\rm sim}, \theta_{\rm sim}$, and $\phi_{\rm sim}$ direction, respectively.
As the same as most of the works on GRMHD simulations of accretion flows, the interval of radial grid points exponentially increases with radius and the grid points in the $\theta_{\rm sim}$ direction concentrates near the equatorial plane of the coordinate system~\cite{gammie2003}.
Because the initial magnetic field is amplified via the magneto-rotational instability (MRI~\cite{balbus1991}), the spatial resolution of the simulation domain can affect the resulting magnetic field strength.
The MRI quantity factor (Q-factor), which evaluates the number of available meshes to resolve the fastest growing mode of MRI~\cite{hawley2011, porth2019}, is $(Q^{r_{\rm sim}}, Q^{\theta_{\rm sim}}, Q^{\phi_{\rm sim}}) \simeq (8.1, 4.9, 18)$. 
Here, in order to evaluate the Q-factors, we analyzed the same region as previous work on non-tilted accretion flows~\cite{porth2019}, except that we extended the region by $\pm 15^{\circ}$ in $\theta_{\rm sim}$-direction, i.e., the $45^{\circ} \le \theta_{\rm sim} \le 135^{\circ}$ to take into account the precession of the disk with initial tilt angle $15^{\circ}$.
The resulting Q-factors are smaller than the required values suggested in a previous work~\cite{hawley2011} ($Q^z \sim 10$ and $Q^{\phi_{\rm sim}} \sim 20$, in the cylindrical coordinates), however, are satisfying the ones proposed by another previous work~\cite{sano2004} ($Q \sim 6$ in the Cartesian coordinates).
Therefore, this simulation would marginally resolve the growth of MRI.

For the analysis of the tilt and precession angles, we follow a similar manner as described in a previous work~\cite{fragile2007}. 
We evaluate the tilt angle $\psi_{\rm tilt}$ and the precession angle $\phi_{\rm prec} (r_{\rm sim}) $ at a certain radius as follows:
\begin{eqnarray}
    \Psi_{\rm tilt} (r_{\rm sim}) &=& \arccos{\qty(\frac{{\vb* J}_{\rm BH}\vdot {\vb* J}_{\rm MHD}(r_{\rm sim})}{\abs{{\vb* J}_{\rm BH}} \abs{{\vb* J}_{\rm MHD}(r_{\rm sim})}})}, \\
    \phi_{\rm prec} (r_{\rm sim}) &=& \arccos{\qty(\frac{{\vb* J}_{\rm BH} \cross {\vb* J}_{\rm MHD}(r_{\rm sim})}{\abs{{\vb* J}_{\rm BH} \cross {\vb* J}_{\rm MHD}(r_{\rm sim})}} \vdot {\vb* e}_{ y})},
\end{eqnarray}
where $e_{\rm y}$ is the unit vector along the y axis, ${\vb* J}_{\rm BH}$ is the dimensionless angular momentum vector of the BH and ${\vb* J}_{\rm MHD}$ is the angular momentum of the MHD plasma in an asymptotically flat space:
\begin{eqnarray}
    {\vb* J}_{\rm BH} &=& a_{*} {\vb* e_{z}}, \\
    {\vb* J}_{\rm MHD} &=& J^{ x}_{\rm MHD} {\vb* e_{ x}} + J^{ y}_{\rm MHD} {\vb* e_{ y}} + J^{ z}_{\rm MHD} {\vb* e_{ z}}.
\end{eqnarray}
We define $J^{ i}_{\rm MHD} ({ i} = { x}, { y}, { z})$ as follows:
\begin{eqnarray}
    \qty(J_{\rm MHD})_{\delta} = \frac{\epsilon_{\alpha \beta \gamma \delta} L^{\alpha \beta} P^{\gamma}}{2 \sqrt{P^{\mu}P_{\mu}}},
\end{eqnarray}
where $L^{\alpha \beta}$ and $P^{\gamma} (r_{\rm sim})$ are the total angular momentum and total four-momentum inside the shell of the width $\Delta r$, respectively, which are described as 
\begin{eqnarray}
     L^{\alpha \beta} (r_{\rm sim}) &=& \int (x^{\alpha} T^{\beta 0} - x^{\beta} T^{\alpha 0} ) \dd[3]x, \\
     P^{\gamma} (r_{\rm sim}) &=& \int T^{\gamma 0} \dd[3]x.
\end{eqnarray}
It will be useful to note that $P^{\gamma}(r_{\rm sim})/\sqrt{P^{\mu} (r_{\rm sim}) P_{\mu} (r_{\rm sim})}$ is the four-velocity of the mass center.

It should be noted that in our simulation an LT precession of the accretion disk occurs with an almost constant precession period after the system evolves into a steadily precessing state, which is in agreement with Ref.~\cite{white2019}. For their $a_* =0.9$ case, the precession rate is $\sim4^\circ$ per 1,000\,$r_{\rm g}/c$, corresponding to a period of $T\sim90$\,years after adopting M87 BH mass. The shorter period in our case (i.e., $T\sim11$\,years that match the inferred period from observation) would be attributed to the final disk size, higher magnitude of BH spin $a_* = 0.9375$, and/or our larger specific heat ratio 13/9, which will result in the rapid wave propagation in the disk for the rigid-body precession~\cite{liu_2002}.
As mentioned above, our simulation marginally resolves MRI. Regarding this, Ref.~\cite{liska2018} raises the caution that if MRI is sufficiently resolved, the disk would expand and the precession would be slowed down. However, the consistency between our simulation and that with a higher resolution performed in Ref.~\cite{white2019} somehow justifies our simulated result. Although there may still be uncertainties regarding whether the simulation time is long enough, a smaller disk would be a preferred setup for obtaining the inferred period from the observations. Indeed, since the M87 disk size is so far poorly constrained, one may adjust the initial disk size to compensate for the disk expansion if the simulation is really under-resolved. Nonetheless, systematic works are warranted to explore broader parameter space and other physical properties in tilted systems, such as the feeding of an outer disk and the magnetic field morphology. Our observation of the jet precessing provides important information and constraints for numerical studies of tilted-disk systems, especially for M87.

\subsection{Alternative origins of the variations in jet position angle}
\label{sec:jet_prec_model}

Here, we discuss the alternative scenarios which could cause the jet PA variations, including binary BH (BBH) systems, instabilities, and disk-jet interactions.

In the BBH system, a precessing jet is developed if the primary BH has an accretion disk that is not co-planar with the binary system orbit. The disk is forced to precess by the effect of the torque from the secondary BH~\cite{abraham2018, britzen2018}. This situation is similar to a tilted disk system. The formed jet from the primary BH will precess as seen in tilted disk simulations. However, we do not have any observational evidence of a BBH system in M87. One of the best candidates for the SMBH system OJ287 has presented quasi-periodic double-peaked optical outbursts that have been interpreted as produced by a secondary BH impacting twice the accretion disk of the primary. In M87, such quasi-periodic outbursts have not been observed yet. If the M87 is a BBH system, we may see the position change of the radio core of the M87. However, from our long-term radio monitoring of the radio core of M87, we do not see such evidence. The horizon-scale images of M87 by the EHT observation have not shown any structure by a secondary BH. From the observational evidence, we think the BBH scenario is not preferred.

In the instabilities, we have a possibility to grow two major types, Kelvin-Helmholtz (KH) and current-driven (CD) kink instabilities during jet propagation. KH instability is excited by the velocity shear which naturally happens at the boundary between the jet and the external medium. A helical mode of KH instability will develop a helical structure inside the jet. However, the existence of a strong magnetic field suppresses the growth of KH instability~\cite{mizuno2007}. Our observed jet region is located jet acceleration and collimation zone. From the jet formation mechanism by the MHD process, in such a region magnetic field is dominated. Thus, KH instability is not suitable for the origin of the precessing jet of M87. Instead of KH instability, CD kink instability will grow in the jet which is excited by the existence of a helical magnetic field. A helical magnetic field is naturally expected from GRMHD simulations of jet formation~\cite{porth2019, cruz2022}. CD kink instability is faster growth in a strongly magnetized region and a developed helically twisted jet structure. The helically twisted structure is advected along the jet while expanding radially~\cite{mizuno2014, singh2016}. 
The growth rate of CD kink instability depends on the magnetic pitch (the ratio of poloidal and toroidal magnetic field) and local Alfven speed (i.e., magnetic field strength). In general, the growth rate of CD kink instability is different with jet radius. However, such a feature of varying amplitude at different locations is not apparent within our selected regions as viewed from Fig.~1 and from the consistency between the variation amplitudes obtained from two analyses using different jet distance ranges, namely, 0.7--3\,mas versus 1.7--3\,mas from the core. Thus, the CD kink instability scenario is also disfavored.

In the disk-jet interaction scenario, the jet structure is affected by the inhomogeneous mass accretion onto a BH (mass injection to the jet). In the MAD phase, mass accretion onto a BH is stopped locally by strong magnetic pressure~\cite{tchek2011}. Disruption of accretion flows will have a certain period. It would be possible to make a quasi-period mass accretion. However, the time scale is roughly several 1,000\,$r_{\rm g}/c$ which is shorter than the observed period. Such local disruption of accretion flows will trigger the excitement of instabilities and produce an asymmetric structure in the jet. They would be the same as the instability scenario.
\pagebreak
\renewcommand{\figurename}{Extended Data Figure}
\setcounter{figure}{0}

\begin{figure}[htbp]
 \begin{center}
 \includegraphics[width=1\textwidth]{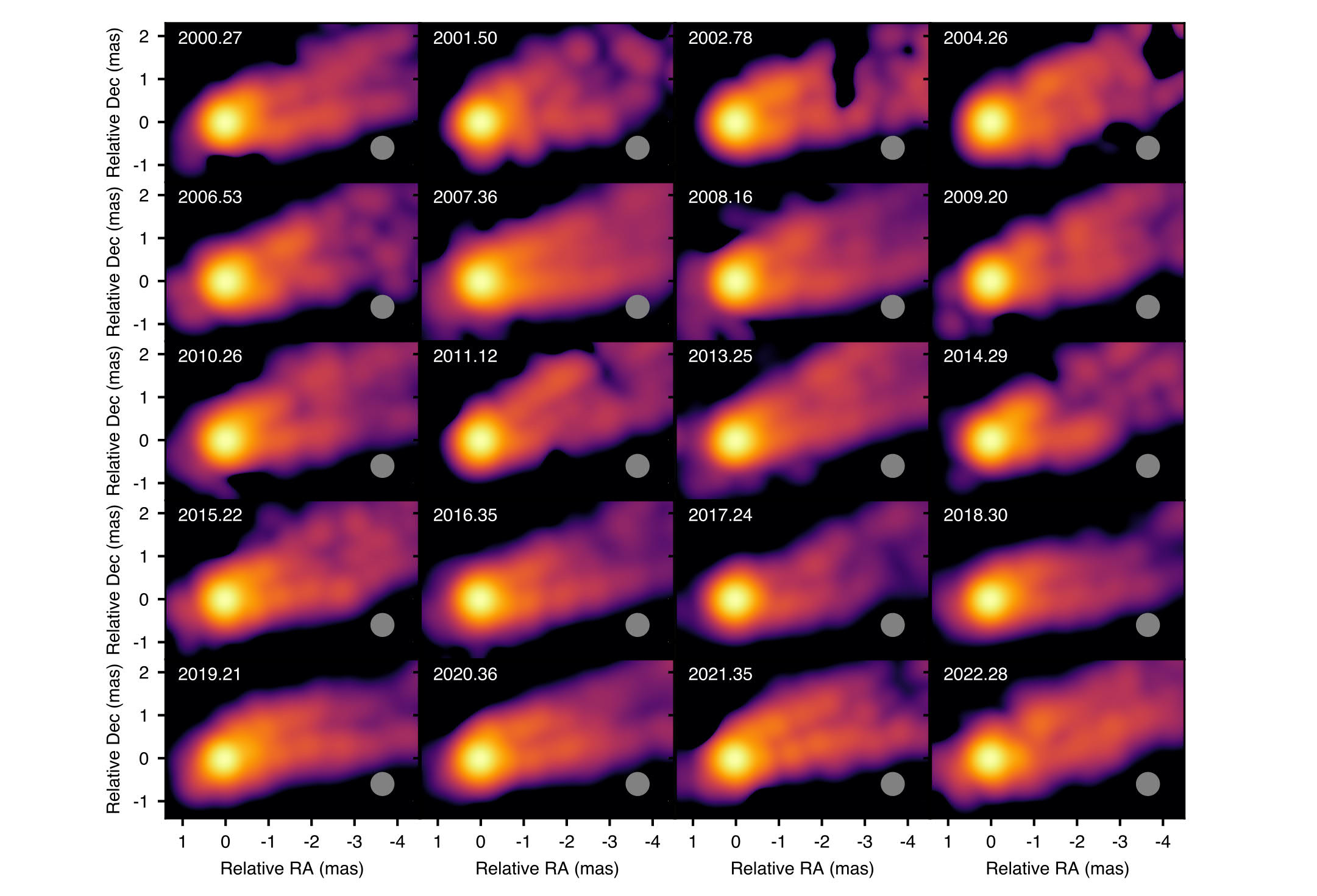}
 \end{center}
 \caption{\textbf{Structural evolution of M87 jet 2000--2022 at Q band.} The images are produced by the yearly stacked EAVN and VLBA data. A common circular restoring beam with FWHM of 0.5\,mas (shown in the bottom-right corner of each panel) is used for all individual images before stacking. The observing year is indicated at the top-left corner.
 }
 \label{fig:ssQ1}
\end{figure}

\pagebreak
\begin{figure}[htbp]
 \begin{center}
 \includegraphics[width=1\textwidth]{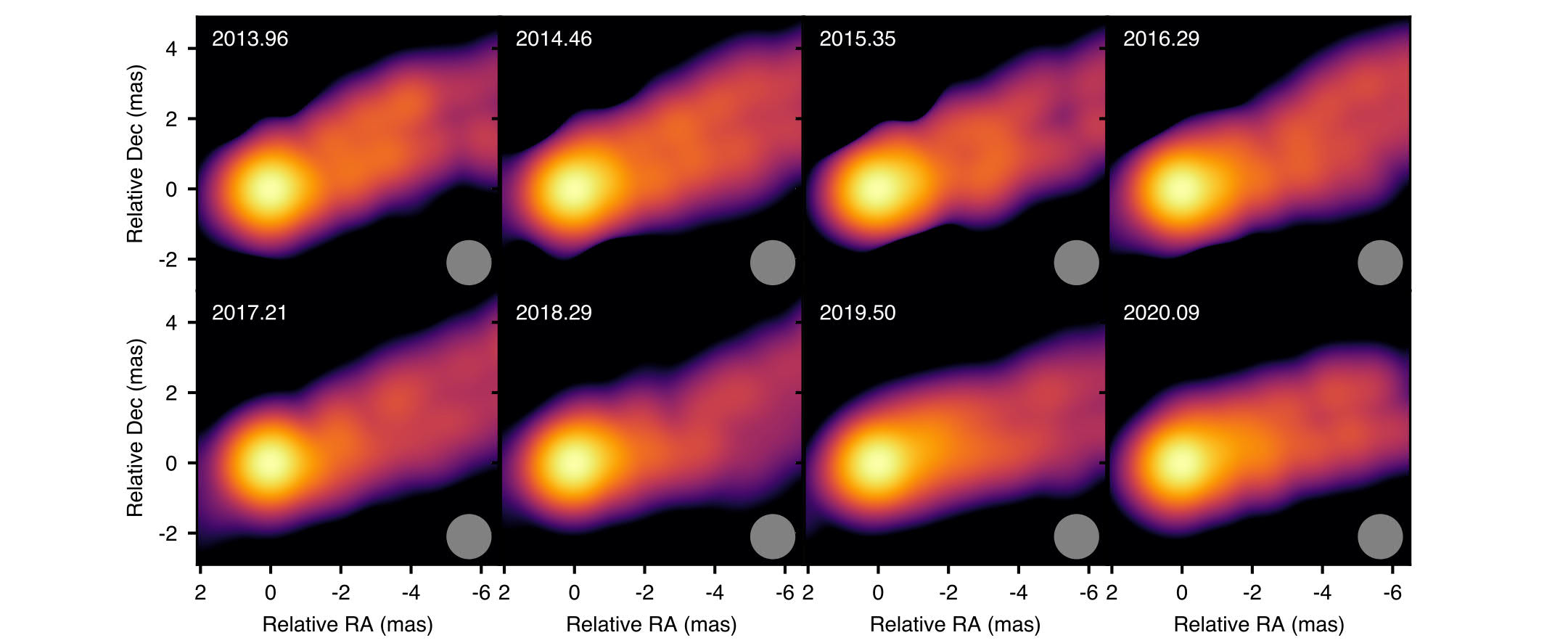}
 \end{center}
 \caption{\textbf{Structural evolution of M87 jet 2013--2020 at K band.} The images are produced by the yearly stacked EAVN and VLBA data. A common circular restoring beam with FWHM of 1.2\,mas (shown in the bottom-right corner of each panel) is used for all individual images before stacking. The observing year is indicated at the top-left corner.
 }
 \label{fig:ssK1}
\end{figure}

\pagebreak
\begin{figure}[htbp]
 \begin{center}
 \includegraphics[width=1\textwidth]{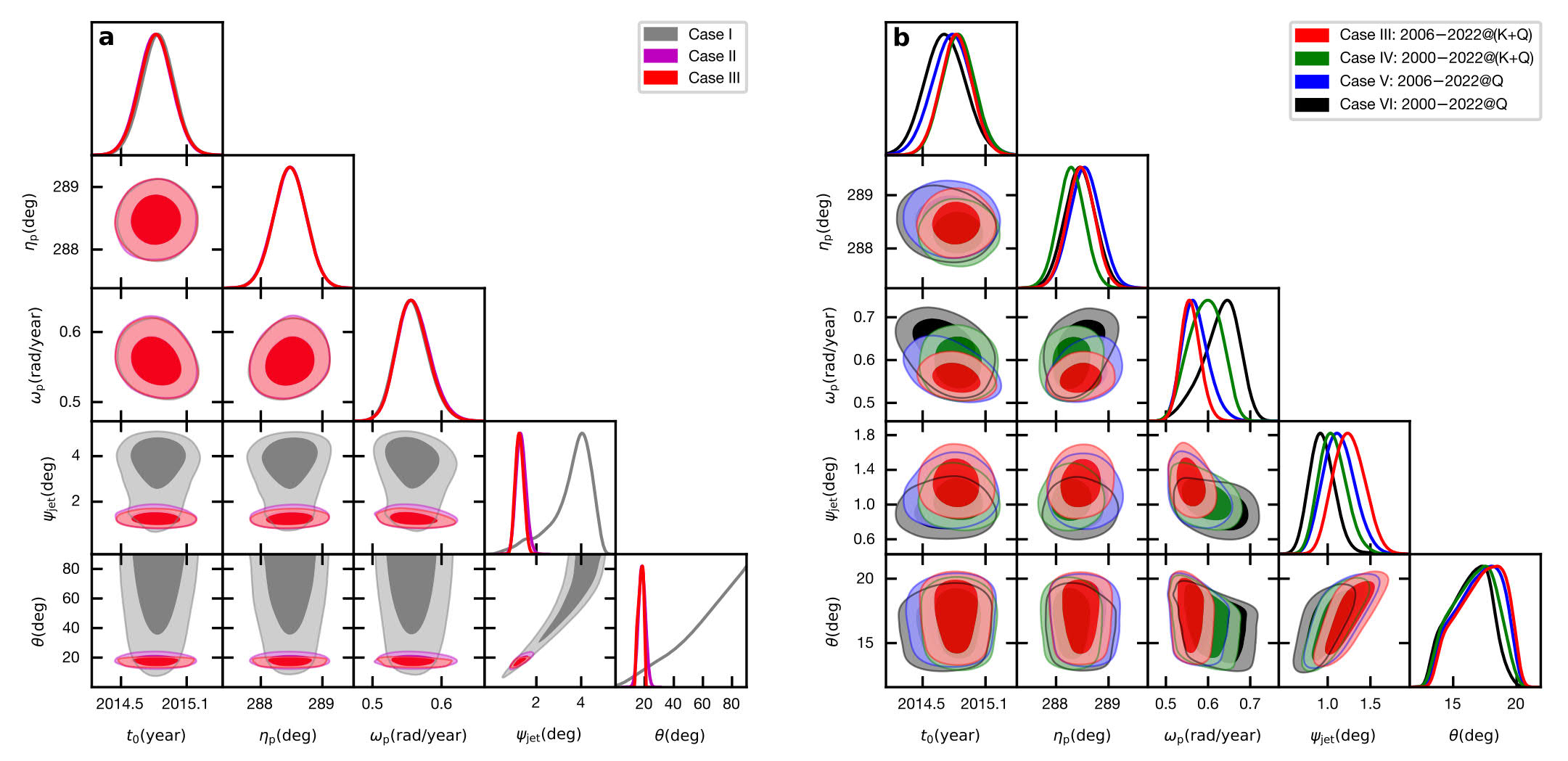}
 \end{center}
 \caption{\textbf{Posterior distributions of precession model parameters in different cases.} \textbf{(a)}: comparison among Case I--III with different constraints. \textbf{(b)}: comparison among Case III--VI with different data sets. The detailed information for each case is described in \EXTTAB{tab:mcmc2} and Methods. The contours correspond to the 68\% and 95\% confidence levels.
 } 
 \label{fig:mcmc1}
\end{figure}

\pagebreak
\begin{figure}[htbp]
 \begin{center}
 \includegraphics[width=1\textwidth]{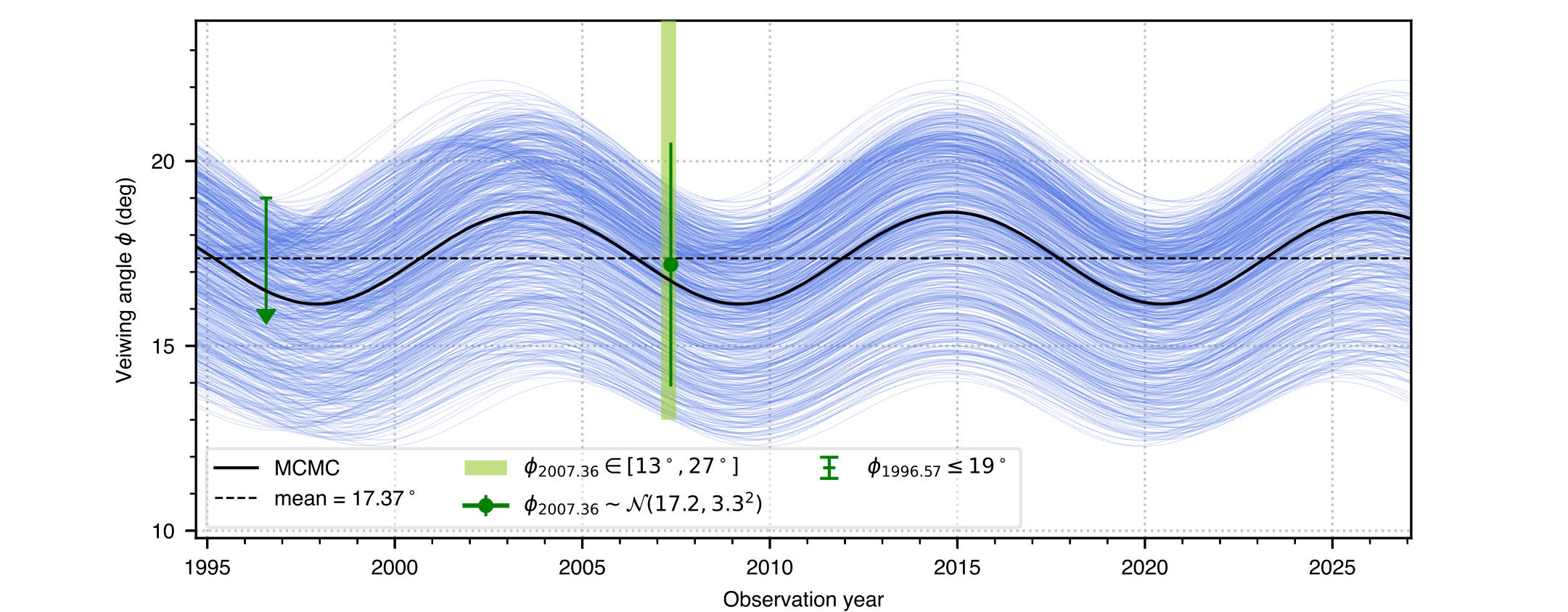}
 \end{center}
 \caption{\textbf{Evolution of the viewing angle $\phi$ as a function of time.} The black thick line is derived from the best-fit precession model parameters. The blue thin lines are plotted by the randomly chosen model parameters derived from the MCMC samples and represent the statistical errors. The constraint of $\phi_{2007.36}\sim\mathcal{N}(17.2,{3.3}^2)$ obtained from Ref.~\cite{mertens2016} is represented by the green dot with an error bar of one standard deviation. The constraints of $\phi_{1996.57}\le 19^\circ$~\cite{biretta1999} and $\phi_{2007.36}\in[13,27]^\circ$~\cite{mertens2016} are indicated with with green arrow and shadow, respectively.
 } 
 \label{fig:mcmc-phi}
\end{figure}
\pagebreak
\renewcommand{\tablename}{Extended Data Table}
\setcounter{table}{0}
\begin{table*}[htbp]
    \begin{tabular*}{1.0\textwidth}{@{\extracolsep{\fill}}lllll}
    \hline
    \hline
Array&Frequency (GHz)&$\Theta^a$\,(mas)&$N_{\rm epoch}^b$&$Y_{\rm ob}^c$\\
    \hline
EAVN$^d$ &22/43  &    0.55/0.63   &119   &2013--2021\\
VLBA      &24/43  &    0.40/0.23 &47    &2006--2018\\
EATING    &22     &    0.27      &4     &2017, 2019, 2020\\
    \hline
    \end{tabular*}
    \caption{\textbf{Summary of the data from the different arrays.} $^a$ Typical angular resolution. $^b$ Number of the epochs. $^c$ Observing years. $^d$ Part of EAVN observations were conducted with only KaVA array. The angular resolution of only KaVA is 1.26\,mas at 22\,GHz and 0.63\,mas at 43\,GHz.}
    \label{tab:array}
\end{table*}

\pagebreak
\begin{table*}[htbp]
    \begin{tabular*}{1.0\textwidth}{@{\extracolsep{\fill}}lccl}
    \hline
    \hline
    Epoch   & Obs. Date&$N_{\rm Ant}^a$ & Stations   \\
    \hline
    a17107a & 2017-04-17& 11 &KaVA$^b$ (no KUS), TMRT$^c$, NSRT$^d$, HIT$^e$, MDC$^f$\\
    a19mk02q& 2019-12-06& 9 &KaVA, NSRT, SRT$^g$\\
    a19mk02r& 2019-12-21& 9 &KaVA, NSRT, BDR$^h$\\
    a2015a  & 2020-01-30& 8 &KaVA, SRT\\
    \hline
    \end{tabular*}
    \caption{\textbf{Antenna information of four EATING observations at 22\,GHz.} $^a$ Number of participated antenna. $^b$ KaVA: Korean VLBI Network (KVN) and VERA, including Mizusawa-20m, Iriki-20m, Ishigaki-20m, Ogasawara-20m telescopes in Japan and Tamna-21m, Ulsan-21m, Yonsei-21m telescopes in Korea. $^c$ Tianma-65m telescope in China. $^d$ Nanshan-26m telescope in China. $^e$ Hitachi-32m telescope in Japan. $^f$ Medicina-32m telescope in Italy. $^g$ Sardinia-64m telescope in Italy. $^h$ Badary-32m telescope in Russia.}
    \label{tab:eating}
\end{table*}

\pagebreak
\begin{table*}[htbp]
    \begin{tabular*}{1.0\textwidth}{@{\extracolsep{\fill}}llll}
    \hline
    \hline
     Parameter       & Type     & Details                   &Unit\\
    \hline
     $t_0$           &  Uniform &  [2012, 2018]             & year    \\
     $\eta_{\rm p}$  &  Uniform &  [280, 295]               & deg    \\
     $\omega_{\rm p}$&  Uniform &  [0, 1]                   & radian/year \\
     $\psi_{\rm jet}$          &  Uniform &  [0, 5]                   & deg    \\
    $\theta$         &  Uniform &  [0, 90]                  & deg    \\
    \hline
    \end{tabular*}
    \caption{\textbf{Common prior distribution for each parameter in Case I -- Case VI.} The specifications for different cases are listed in \EXTTAB{tab:mcmc2}.}
    \label{tab:mcmc1}
\end{table*}

\clearpage
\begin{table*}[htbp]
    \begin{tabular*}{1.0\textwidth}{@{\extracolsep{\fill}}lll}
    \hline
    \hline
     Label       &  Data & Additional constraints                   \\
    \hline
    Case I & 2006--2022 at K/Q bands&no additional constraints\\
    Case II & same as above &$\phi_{2007.36}\sim\mathcal{N}(17.2,{3.3}^2)$~\cite{mertens2016}            \\
    Case III & same as above& $\phi_{2007.36}\sim\mathcal{N}(17.2,{3.3}^2)$, $\phi_{2007.36}\in[13,27]^\circ$~\cite{mertens2016}, $\phi_{1996.57}\le 19^\circ$~\cite{biretta1999}\\
    Case IV & 2000--2022 at K/Q bands& same as above\\
    Case V & 2006--2022 at Q band& same as above\\
    Case VI & 2000--2022 at Q band& same as above\\
\hline
    \end{tabular*}
    \caption{\textbf{Detailed specifications for Case I -- Case VI.} $\phi_{2007.36}\sim\mathcal{N}(17.2,{3.3}^2)$ indicates the Gaussian distribution for $\phi$ at $t=2007.36$\,years based on Ref.~\cite{mertens2016}. The corresponding MCMC fitting results are compared in the \EXTFIG{fig:mcmc1} and listed in \EXTTAB{tab:mcmc3}.}
    \label{tab:mcmc2}
\end{table*}

\clearpage
\begin{table*}[htbp]
\resizebox{\textwidth}{!}{
    \begin{tabular}{cccccccc}
    \hline
    \hline
         & $t_0$& $\eta_{\rm p}$ ($^\circ$)      &$\omega_{\rm p}$ (radian/year)  &  $\psi_{\rm jet}$ ($^\circ$)&$\theta$ ($^\circ$)   & $T_{\rm prec}^{\rm jet}$(year) & $\hat{\chi}^2$\\
    \hline       
Case I	& $2014.84\pm0.15$ &	$288.47\pm0.27$ &	$0.56\pm0.02$	& $3.56\pm0.93$ &	$62.71\pm20.54$& $11.27\pm0.48$ &	N/A$^a$	\\
Case II	& $2014.82\pm0.15$ &	$288.47\pm0.26$ &	$0.56\pm0.02$	& $1.32\pm0.22$ &	$18.14\pm2.40$ & $11.22\pm0.48$ &	1.25	\\
Case III& $2014.82\pm0.15$ &	$288.47\pm0.27$ &	$0.56\pm0.02$	& $1.25\pm0.18$ &	$17.21\pm1.74$ & $11.24\pm0.47$ &	1.25	\\
Case IV	& $2014.84\pm0.16$ &	$288.29\pm0.26$ &	$0.59\pm0.04$	& $1.06\pm0.16$ &	$16.63\pm1.68$ & $10.58\pm0.68$ &	1.98	\\
Case V	& $2014.78\pm0.19$ &	$288.55\pm0.30$ &	$0.57\pm0.03$	& $1.13\pm0.18$ &	$17.01\pm1.73$ & $11.01\pm0.61$ &	1.53	\\
Case VI	& $2014.72\pm0.19$ &	$288.45\pm0.30$ &	$0.63\pm0.04$	& $0.94\pm0.15$ &	$16.31\pm1.57$ & $9.99\pm0.70$ &	2.32	\\

    \hline
    \end{tabular}
    }
    \caption{\textbf{MCMC fitting results for Case I -- Case VI.} The last column is the reduced $\hat{\chi}^2$ value calculated with the best-fit model parameters (Equation~\eqref{eq:reduced-chi2}). We adopt Case III as the final fitting results as shown in Table~\ref{tab:fitp}. $^a$ For Case I, $\hat{\chi}^2$ is not applicable since there is a degeneracy between $\psi_{\rm jet}$ and $\theta$. The values correspond to the means of the MCMC samples with standard deviations.}
    \label{tab:mcmc3}
\end{table*}

\pagebreak
\begin{table*}[htbp]
    \begin{tabular*}{1.0\textwidth}{@{\extracolsep{\fill}}llllllll}
    \hline
    \hline
     Year  &  2017   & 2018   & 2019   & 2020   & 2021   & 2022   & 2023\\
    \hline
     $\phi$ (deg)   & $17.7\pm1.8$   & $17.0\pm1.7$   & $16.4\pm1.7$   & $16.0\pm1.6$   & $16.0\pm1.6$   & $16.4\pm1.6$   & $17.0\pm1.7$\\  
    \hline
    \end{tabular*}
    \caption{\textbf{Jet viewing angle at some selected years.} Errors are standard deviation.}
    \label{tab:mcmc-phi}
\end{table*}

\end{methods}
\pagebreak
\section*{References}

\begin{thebibliography}{10}
\expandafter\ifx\csname url\endcsname\relax
  \def\url#1{\texttt{#1}}\fi
\expandafter\ifx\csname urlprefix\endcsname\relax\def\urlprefix{URL }\fi
\providecommand{\bibinfo}[2]{#2}
\providecommand{\eprint}[2][]{\url{#2}}

\bibitem{junor1999}
\bibinfo{author}{{Junor}, W.}, \bibinfo{author}{{Biretta}, J.~A.} \& \bibinfo{author}{{Livio}, M.}
\newblock \bibinfo{title}{{Formation of the radio jet in M87 at 100 Schwarzschild radii from the central black hole}}.
\newblock \emph{\bibinfo{journal}{\nat}} \textbf{\bibinfo{volume}{401}}, \bibinfo{pages}{891--892} (\bibinfo{year}{1999}).

\bibitem{hada2011n}
\bibinfo{author}{{Hada}, K.} \emph{et~al.}
\newblock \bibinfo{title}{{An origin of the radio jet in M87 at the location of the central black hole}}.
\newblock \emph{\bibinfo{journal}{\nat}} \textbf{\bibinfo{volume}{477}}, \bibinfo{pages}{185--187} (\bibinfo{year}{2011}).

\bibitem{walker2018}
\bibinfo{author}{{Walker}, R.~C.}, \bibinfo{author}{{Hardee}, P.~E.}, \bibinfo{author}{{Davies}, F.~B.}, \bibinfo{author}{{Ly}, C.} \& \bibinfo{author}{{Junor}, W.}
\newblock \bibinfo{title}{{The Structure and Dynamics of the Subparsec Jet in M87 Based on 50 VLBA Observations over 17 Years at 43 GHz}}.
\newblock \emph{\bibinfo{journal}{\apj}} \textbf{\bibinfo{volume}{855}}, \bibinfo{pages}{128} (\bibinfo{year}{2018}).

\bibitem{lu2023}
\bibinfo{author}{{Lu}, R.} \emph{et~al.}
\newblock \bibinfo{title}{{A ring-like accretion structure in M87 connecting its black hole and jet}}.
\newblock \emph{\bibinfo{journal}{\nat}} \textbf{\bibinfo{volume}{616}}, \bibinfo{pages}{686--690} (\bibinfo{year}{2023}).

\bibitem{eht12019}
\bibinfo{author}{{Event Horizon Telescope Collaboration}}.
\newblock \bibinfo{title}{{First M87 Event Horizon Telescope Results. I. The Shadow of the Supermassive Black Hole}}.
\newblock \emph{\bibinfo{journal}{\apjl}} \textbf{\bibinfo{volume}{875}}, \bibinfo{pages}{L1} (\bibinfo{year}{2019}).

\bibitem{cui2021}
\bibinfo{author}{{Cui}, Y.-Z.} \emph{et~al.}
\newblock \bibinfo{title}{{East Asian VLBI Network observations of active galactic nuclei jets: imaging with KaVA+Tianma+Nanshan}}.
\newblock \emph{\bibinfo{journal}{\raa}} \textbf{\bibinfo{volume}{21}}, \bibinfo{pages}{205} (\bibinfo{year}{2021}).

\bibitem{giovannini2023}
\bibinfo{author}{{Giovannini}, G.} \emph{et~al.}
\newblock \bibinfo{title}{{The Past and Future of East Asia to Italy: Nearly Global VLBI}}.
\newblock \emph{\bibinfo{journal}{Galaxies}} \textbf{\bibinfo{volume}{11}}, \bibinfo{pages}{49} (\bibinfo{year}{2023}).

\bibitem{hada2016gmva}
\bibinfo{author}{{Hada}, K.} \emph{et~al.}
\newblock \bibinfo{title}{{High-sensitivity 86 GHz (3.5 mm) VLBI Observations of M87: Deep Imaging of the Jet Base at a Resolution of 10 Schwarzschild Radii}}.
\newblock \emph{\bibinfo{journal}{\apj}} \textbf{\bibinfo{volume}{817}}, \bibinfo{pages}{131} (\bibinfo{year}{2016}).

\bibitem{hada2014}
\bibinfo{author}{{Hada}, K.} \emph{et~al.}
\newblock \bibinfo{title}{{A Strong Radio Brightening at the Jet Base of M 87 during the Elevated Very High Energy Gamma-Ray State in 2012}}.
\newblock \emph{\bibinfo{journal}{\apj}} \textbf{\bibinfo{volume}{788}}, \bibinfo{pages}{165} (\bibinfo{year}{2014}).

\bibitem{park2019v}
\bibinfo{author}{{Park}, J.} \emph{et~al.}
\newblock \bibinfo{title}{{Kinematics of the M87 Jet in the Collimation Zone: Gradual Acceleration and Velocity Stratification}}.
\newblock \emph{\bibinfo{journal}{\apj}} \textbf{\bibinfo{volume}{887}}, \bibinfo{pages}{147} (\bibinfo{year}{2019}).

\bibitem{ro2023}
\bibinfo{author}{{Ro}, H.} \emph{et~al.}
\newblock \bibinfo{title}{{Transverse Oscillations of the M87 Jet Revealed by KaVA Observations}}.
\newblock \emph{\bibinfo{journal}{Galaxies}} \textbf{\bibinfo{volume}{11}}, \bibinfo{pages}{33} (\bibinfo{year}{2023}).

\bibitem{caproni2004a}
\bibinfo{author}{{Caproni}, A.} \& \bibinfo{author}{{Abraham}, Z.}
\newblock \bibinfo{title}{{Precession in the Inner Jet of 3C 345}}.
\newblock \emph{\bibinfo{journal}{\apj}} \textbf{\bibinfo{volume}{602}}, \bibinfo{pages}{625--634} (\bibinfo{year}{2004}).

\bibitem{mizuno2007}
\bibinfo{author}{{Mizuno}, Y.}, \bibinfo{author}{{Hardee}, P.} \& \bibinfo{author}{{Nishikawa}, K.-I.}
\newblock \bibinfo{title}{{Three-dimensional Relativistic Magnetohydrodynamic Simulations of Magnetized Spine-Sheath Relativistic Jets}}.
\newblock \emph{\bibinfo{journal}{\apj}} \textbf{\bibinfo{volume}{662}}, \bibinfo{pages}{835--850} (\bibinfo{year}{2007}).

\bibitem{mizuno2012}
\bibinfo{author}{{Mizuno}, Y.}, \bibinfo{author}{{Lyubarsky}, Y.}, \bibinfo{author}{{Nishikawa}, K.-I.} \& \bibinfo{author}{{Hardee}, P.~E.}
\newblock \bibinfo{title}{{Three-dimensional Relativistic Magnetohydrodynamic Simulations of Current-driven Instability. III. Rotating Relativistic Jets}}.
\newblock \emph{\bibinfo{journal}{\apj}} \textbf{\bibinfo{volume}{757}}, \bibinfo{pages}{16} (\bibinfo{year}{2012}).

\bibitem{lense1918}
\bibinfo{author}{{Lense}, J.} \& \bibinfo{author}{{Thirring}, H.}
\newblock \bibinfo{title}{{{\"U}ber den Einflu{\ss} der Eigenrotation der Zentralk{\"o}rper auf die Bewegung der Planeten und Monde nach der Einsteinschen Gravitationstheorie}}.
\newblock \emph{\bibinfo{journal}{Physikalische Zeitschrift}} \textbf{\bibinfo{volume}{19}}, \bibinfo{pages}{156--163} (\bibinfo{year}{1918}).

\bibitem{fragile2007}
\bibinfo{author}{{Fragile}, P.~C.}, \bibinfo{author}{{Blaes}, O.~M.}, \bibinfo{author}{{Anninos}, P.} \& \bibinfo{author}{{Salmonson}, J.~D.}
\newblock \bibinfo{title}{{Global General Relativistic Magnetohydrodynamic Simulation of a Tilted Black Hole Accretion Disk}}.
\newblock \emph{\bibinfo{journal}{\apj}} \textbf{\bibinfo{volume}{668}}, \bibinfo{pages}{417--429} (\bibinfo{year}{2007}).

\bibitem{mcKinney2013}
\bibinfo{author}{{McKinney}, J.~C.}, \bibinfo{author}{{Tchekhovskoy}, A.} \& \bibinfo{author}{{Blandford}, R.~D.}
\newblock \bibinfo{title}{{Alignment of Magnetized Accretion Disks and Relativistic Jets with Spinning Black Holes}}.
\newblock \emph{\bibinfo{journal}{Science}} \textbf{\bibinfo{volume}{339}}, \bibinfo{pages}{49--52} (\bibinfo{year}{2013}).

\bibitem{liska2018}
\bibinfo{author}{{Liska}, M.} \emph{et~al.}
\newblock \bibinfo{title}{{Formation of precessing jets by tilted black hole discs in 3D general relativistic MHD simulations}}.
\newblock \emph{\bibinfo{journal}{\mnras}} \textbf{\bibinfo{volume}{474}}, \bibinfo{pages}{L81--L85} (\bibinfo{year}{2018}).

\bibitem{white2019}
\bibinfo{author}{{White}, C.~J.}, \bibinfo{author}{{Quataert}, E.} \& \bibinfo{author}{{Blaes}, O.}
\newblock \bibinfo{title}{{Tilted Disks around Black Holes: A Numerical Parameter Survey for Spin and Inclination Angle}}.
\newblock \emph{\bibinfo{journal}{\apj}} \textbf{\bibinfo{volume}{878}}, \bibinfo{pages}{51} (\bibinfo{year}{2019}).

\bibitem{chatterjee2020}
\bibinfo{author}{{Chatterjee}, K.} \emph{et~al.}
\newblock \bibinfo{title}{{Observational signatures of disc and jet misalignment in images of accreting black holes}}.
\newblock \emph{\bibinfo{journal}{\mnras}} \textbf{\bibinfo{volume}{499}}, \bibinfo{pages}{362--378} (\bibinfo{year}{2020}).

\bibitem{ressler2023}
\bibinfo{author}{{Ressler}, S.~M.}, \bibinfo{author}{{White}, C.~J.} \& \bibinfo{author}{{Quataert}, E.}
\newblock \bibinfo{title}{{Wind-fed GRMHD simulations of Sagittarius A*: tilt and alignment of jets and accretion discs, electron thermodynamics, and multiscale modelling of the rotation measure}}.
\newblock \emph{\bibinfo{journal}{\mnras}} \textbf{\bibinfo{volume}{521}}, \bibinfo{pages}{4277--4298} (\bibinfo{year}{2023}).

\bibitem{takahashi2016}
\bibinfo{author}{{Takahashi}, H.~R.}, \bibinfo{author}{{Ohsuga}, K.}, \bibinfo{author}{{Kawashima}, T.} \& \bibinfo{author}{{Sekiguchi}, Y.}
\newblock \bibinfo{title}{{Formation of Overheated Regions and Truncated Disks around Black Holes: Three-dimensional General Relativistic Radiation-magnetohydrodynamics Simulations}}.
\newblock \emph{\bibinfo{journal}{\apj}} \textbf{\bibinfo{volume}{826}}, \bibinfo{pages}{23} (\bibinfo{year}{2016}).

\bibitem{kawashima2023}
\bibinfo{author}{{Kawashima}, T.}, \bibinfo{author}{{Ohsuga}, K.} \& \bibinfo{author}{{Takahashi}, H.~R.}
\newblock \bibinfo{title}{{RAIKOU: A General Relativistic, Multiwavelength Radiative Transfer Code}}.
\newblock \emph{\bibinfo{journal}{\apj}} \textbf{\bibinfo{volume}{949}}, \bibinfo{pages}{101} (\bibinfo{year}{2023}).

\bibitem{liu_2002}
\bibinfo{author}{{Liu}, S.} \& \bibinfo{author}{{Melia}, F.}
\newblock \bibinfo{title}{{An Accretion-induced X-Ray Flare in Sagittarius A*}}.
\newblock \emph{\bibinfo{journal}{\apjl}} \textbf{\bibinfo{volume}{566}}, \bibinfo{pages}{L77--L80} (\bibinfo{year}{2002}).

\bibitem{blandford1977}
\bibinfo{author}{{Blandford}, R.~D.} \& \bibinfo{author}{{Znajek}, R.~L.}
\newblock \bibinfo{title}{{Electromagnetic extraction of energy from Kerr black holes.}}
\newblock \emph{\bibinfo{journal}{\mnras}} \textbf{\bibinfo{volume}{179}}, \bibinfo{pages}{433--456} (\bibinfo{year}{1977}).

\bibitem{wilkins1972}
\bibinfo{author}{{Wilkins}, D.~C.}
\newblock \bibinfo{title}{{Bound Geodesics in the Kerr Metric}}.
\newblock \emph{\bibinfo{journal}{\prd}} \textbf{\bibinfo{volume}{5}}, \bibinfo{pages}{814--822} (\bibinfo{year}{1972}).

\bibitem{caproni2007}
\bibinfo{author}{{Caproni}, A.}, \bibinfo{author}{{Abraham}, Z.}, \bibinfo{author}{{Livio}, M.} \& \bibinfo{author}{{Mosquera Cuesta}, H.~J.}
\newblock \bibinfo{title}{{Is the Bardeen-Petterson effect responsible for the warping and precession in NGC4258?}}
\newblock \emph{\bibinfo{journal}{\mnras}} \textbf{\bibinfo{volume}{379}}, \bibinfo{pages}{135--142} (\bibinfo{year}{2007}).

\bibitem{scheuer1996}
\bibinfo{author}{{Scheuer}, P.~A.~G.} \& \bibinfo{author}{{Feiler}, R.}
\newblock \bibinfo{title}{{The realignment of a black hole misaligned with its accretion disc}}.
\newblock \emph{\bibinfo{journal}{\mnras}} \textbf{\bibinfo{volume}{282}}, \bibinfo{pages}{291--294} (\bibinfo{year}{1996}).

\bibitem{natarajan1999}
\bibinfo{author}{{Natarajan}, P.} \& \bibinfo{author}{{Armitage}, P.~J.}
\newblock \bibinfo{title}{{Warped discs and the directional stability of jets in active galactic nuclei}}.
\newblock \emph{\bibinfo{journal}{\mnras}} \textbf{\bibinfo{volume}{309}}, \bibinfo{pages}{961--968} (\bibinfo{year}{1999}).

\bibitem{wielgus2020}
\bibinfo{author}{{Wielgus}, M.} \& \bibinfo{author}{{Event Horizon Telescope Collaboration}}.
\newblock \bibinfo{title}{{Monitoring the Morphology of M87* in 2009-2017 with the Event Horizon Telescope}}.
\newblock \emph{\bibinfo{journal}{\apj}} \textbf{\bibinfo{volume}{901}}, \bibinfo{pages}{67} (\bibinfo{year}{2020}).

\bibitem{greisen2003}
\bibinfo{author}{{Greisen}, E.~W.}
\newblock \bibinfo{title}{{AIPS, the VLA, and the VLBA}}.
\newblock In \bibinfo{editor}{{Heck}, A.} (ed.) \emph{\bibinfo{booktitle}{Information Handling in Astronomy - Historical Vistas}}, vol. \bibinfo{volume}{285} of \emph{\bibinfo{series}{Astrophysics and Space Science Library}}, \bibinfo{pages}{109} (\bibinfo{year}{2003}).

\bibitem{hogbom1974}
\bibinfo{author}{{H{\"o}gbom}, J.~A.}
\newblock \bibinfo{title}{{Aperture Synthesis with a Non-Regular Distribution of Interferometer Baselines}}.
\newblock \emph{\bibinfo{journal}{\aaps}} \textbf{\bibinfo{volume}{15}}, \bibinfo{pages}{417--426} (\bibinfo{year}{1974}).

\bibitem{shepherd1994}
\bibinfo{author}{{Shepherd}, M.~C.}, \bibinfo{author}{{Pearson}, T.~J.} \& \bibinfo{author}{{Taylor}, G.~B.}
\newblock \bibinfo{title}{{DIFMAP: an interactive program for synthesis imaging.}}
\newblock In \emph{\bibinfo{booktitle}{\baas}}, vol.~\bibinfo{volume}{26}, \bibinfo{pages}{987--989} (\bibinfo{year}{1994}).

\bibitem{ly2004}
\bibinfo{author}{{Ly}, C.}, \bibinfo{author}{{Walker}, R.~C.} \& \bibinfo{author}{{Wrobel}, J.~M.}
\newblock \bibinfo{title}{{An Attempt to Probe the Radio Jet Collimation Regions in NGC 4278, NGC 4374 (M84), and NGC 6166}}.
\newblock \emph{\bibinfo{journal}{\aj}} \textbf{\bibinfo{volume}{127}}, \bibinfo{pages}{119--124} (\bibinfo{year}{2004}).

\bibitem{ly2007}
\bibinfo{author}{{Ly}, C.}, \bibinfo{author}{{Walker}, R.~C.} \& \bibinfo{author}{{Junor}, W.}
\newblock \bibinfo{title}{{High-Frequency VLBI Imaging of the Jet Base of M87}}.
\newblock \emph{\bibinfo{journal}{\apj}} \textbf{\bibinfo{volume}{660}}, \bibinfo{pages}{200--205} (\bibinfo{year}{2007}).

\bibitem{mertens2016}
\bibinfo{author}{{Mertens}, F.}, \bibinfo{author}{{Lobanov}, A.~P.}, \bibinfo{author}{{Walker}, R.~C.} \& \bibinfo{author}{{Hardee}, P.~E.}
\newblock \bibinfo{title}{{Kinematics of the jet in M 87 on scales of 100-1000 Schwarzschild radii}}.
\newblock \emph{\bibinfo{journal}{\aap}} \textbf{\bibinfo{volume}{595}}, \bibinfo{pages}{A54} (\bibinfo{year}{2016}).

\bibitem{hada2017galax}
\bibinfo{author}{{Hada}, K.}
\newblock \bibinfo{title}{{The Structure and Propagation of the Misaligned Jet M87}}.
\newblock \emph{\bibinfo{journal}{Galaxies}} \textbf{\bibinfo{volume}{5}}, \bibinfo{pages}{2} (\bibinfo{year}{2017}).

\bibitem{ferrari1998}
\bibinfo{author}{{Ferrari}, A.}
\newblock \bibinfo{title}{{Modeling Extragalactic Jets}}.
\newblock \emph{\bibinfo{journal}{\araa}} \textbf{\bibinfo{volume}{36}}, \bibinfo{pages}{539--598} (\bibinfo{year}{1998}).

\bibitem{kino2004}
\bibinfo{author}{{Kino}, M.} \& \bibinfo{author}{{Takahara}, F.}
\newblock \bibinfo{title}{{Constraints on the energetics and plasma composition of relativistic jets in FR II sources}}.
\newblock \emph{\bibinfo{journal}{\mnras}} \textbf{\bibinfo{volume}{349}}, \bibinfo{pages}{336--346} (\bibinfo{year}{2004}).

\bibitem{narayan1995}
\bibinfo{author}{{Narayan}, R.}, \bibinfo{author}{{Yi}, I.} \& \bibinfo{author}{{Mahadevan}, R.}
\newblock \bibinfo{title}{{Explaining the spectrum of Sagittarius A$^{*}$ with a model of an accreting black hole}}.
\newblock \emph{\bibinfo{journal}{\nat}} \textbf{\bibinfo{volume}{374}}, \bibinfo{pages}{623--625} (\bibinfo{year}{1995}).

\bibitem{chatterjee2022}
\bibinfo{author}{{Chatterjee}, K.} \& \bibinfo{author}{{Narayan}, R.}
\newblock \bibinfo{title}{{Flux Eruption Events Drive Angular Momentum Transport in Magnetically Arrested Accretion Flows}}.
\newblock \emph{\bibinfo{journal}{\apj}} \textbf{\bibinfo{volume}{941}}, \bibinfo{pages}{30} (\bibinfo{year}{2022}).

\bibitem{hjellming1981}
\bibinfo{author}{{Hjellming}, R.~M.} \& \bibinfo{author}{{Johnston}, K.~J.}
\newblock \bibinfo{title}{{An analysis of the proper motions of SS 433 radio jets.}}
\newblock \emph{\bibinfo{journal}{\apjl}} \textbf{\bibinfo{volume}{246}}, \bibinfo{pages}{L141--L145} (\bibinfo{year}{1981}).

\bibitem{mcKinney2006}
\bibinfo{author}{{McKinney}, J.~C.}
\newblock \bibinfo{title}{{General relativistic magnetohydrodynamic simulations of the jet formation and large-scale propagation from black hole accretion systems}}.
\newblock \emph{\bibinfo{journal}{\mnras}} \textbf{\bibinfo{volume}{368}}, \bibinfo{pages}{1561--1582} (\bibinfo{year}{2006}).

\bibitem{biretta1999}
\bibinfo{author}{{Biretta}, J.~A.}, \bibinfo{author}{{Sparks}, W.~B.} \& \bibinfo{author}{{Macchetto}, F.}
\newblock \bibinfo{title}{{Hubble Space Telescope Observations of Superluminal Motion in the M87 Jet}}.
\newblock \emph{\bibinfo{journal}{\apj}} \textbf{\bibinfo{volume}{520}}, \bibinfo{pages}{621--626} (\bibinfo{year}{1999}).

\bibitem{foreman2013}
\bibinfo{author}{{Foreman-Mackey}, D.}, \bibinfo{author}{{Hogg}, D.~W.}, \bibinfo{author}{{Lang}, D.} \& \bibinfo{author}{{Goodman}, J.}
\newblock \bibinfo{title}{{emcee: The MCMC Hammer}}.
\newblock \emph{\bibinfo{journal}{\pasp}} \textbf{\bibinfo{volume}{125}}, \bibinfo{pages}{306--312} (\bibinfo{year}{2013}).

\bibitem{gammie2004}
\bibinfo{author}{{Gammie}, C.~F.}, \bibinfo{author}{{Shapiro}, S.~L.} \& \bibinfo{author}{{McKinney}, J.~C.}
\newblock \bibinfo{title}{{Black Hole Spin Evolution}}.
\newblock \emph{\bibinfo{journal}{\apj}} \textbf{\bibinfo{volume}{602}}, \bibinfo{pages}{312--319} (\bibinfo{year}{2004}).

\bibitem{narayan2012}
\bibinfo{author}{{Narayan}, R.}, \bibinfo{author}{{S{\"A} dowski}, A.}, \bibinfo{author}{{Penna}, R.~F.} \& \bibinfo{author}{{Kulkarni}, A.~K.}
\newblock \bibinfo{title}{{GRMHD simulations of magnetized advection-dominated accretion on a non-spinning black hole: role of outflows}}.
\newblock \emph{\bibinfo{journal}{\mnras}} \textbf{\bibinfo{volume}{426}}, \bibinfo{pages}{3241--3259} (\bibinfo{year}{2012}).

\bibitem{porth2019}
\bibinfo{author}{{Porth}, O.} \emph{et~al.}
\newblock \bibinfo{title}{{The Event Horizon General Relativistic Magnetohydrodynamic Code Comparison Project}}.
\newblock \emph{\bibinfo{journal}{\apjs}} \textbf{\bibinfo{volume}{243}}, \bibinfo{pages}{26} (\bibinfo{year}{2019}).

\bibitem{tchek2011}
\bibinfo{author}{{Tchekhovskoy}, A.}, \bibinfo{author}{{Narayan}, R.} \& \bibinfo{author}{{McKinney}, J.~C.}
\newblock \bibinfo{title}{{Efficient generation of jets from magnetically arrested accretion on a rapidly spinning black hole}}.
\newblock \emph{\bibinfo{journal}{\mnras}} \textbf{\bibinfo{volume}{418}}, \bibinfo{pages}{L79--L83} (\bibinfo{year}{2011}).

\bibitem{narayan2022}
\bibinfo{author}{{Narayan}, R.}, \bibinfo{author}{{Chael}, A.}, \bibinfo{author}{{Chatterjee}, K.}, \bibinfo{author}{{Ricarte}, A.} \& \bibinfo{author}{{Curd}, B.}
\newblock \bibinfo{title}{{Jets in magnetically arrested hot accretion flows: geometry, power, and black hole spin-down}}.
\newblock \emph{\bibinfo{journal}{\mnras}} \textbf{\bibinfo{volume}{511}}, \bibinfo{pages}{3795--3813} (\bibinfo{year}{2022}).

\bibitem{white2020}
\bibinfo{author}{{White}, C.~J.}, \bibinfo{author}{{Quataert}, E.} \& \bibinfo{author}{{Gammie}, C.~F.}
\newblock \bibinfo{title}{{The Structure of Radiatively Inefficient Black Hole Accretion Flows}}.
\newblock \emph{\bibinfo{journal}{\apj}} \textbf{\bibinfo{volume}{891}}, \bibinfo{pages}{63} (\bibinfo{year}{2020}).

\bibitem{anantua2020}
\bibinfo{author}{{Anantua}, R.}, \bibinfo{author}{{Ressler}, S.} \& \bibinfo{author}{{Quataert}, E.}
\newblock \bibinfo{title}{{On the comparison of AGN with GRMHD simulations: I. Sgr A*}}.
\newblock \emph{\bibinfo{journal}{\mnras}} \textbf{\bibinfo{volume}{493}}, \bibinfo{pages}{1404--1418} (\bibinfo{year}{2020}).

\bibitem{gammie2003}
\bibinfo{author}{{Gammie}, C.~F.}, \bibinfo{author}{{McKinney}, J.~C.} \& \bibinfo{author}{{T{\'o}th}, G.}
\newblock \bibinfo{title}{{HARM: A Numerical Scheme for General Relativistic Magnetohydrodynamics}}.
\newblock \emph{\bibinfo{journal}{\apj}} \textbf{\bibinfo{volume}{589}}, \bibinfo{pages}{444--457} (\bibinfo{year}{2003}).

\bibitem{balbus1991}
\bibinfo{author}{{Balbus}, S.~A.} \& \bibinfo{author}{{Hawley}, J.~F.}
\newblock \bibinfo{title}{{A Powerful Local Shear Instability in Weakly Magnetized Disks. I. Linear Analysis}}.
\newblock \emph{\bibinfo{journal}{\apj}} \textbf{\bibinfo{volume}{376}}, \bibinfo{pages}{214} (\bibinfo{year}{1991}).

\bibitem{hawley2011}
\bibinfo{author}{{Hawley}, J.~F.}, \bibinfo{author}{{Guan}, X.} \& \bibinfo{author}{{Krolik}, J.~H.}
\newblock \bibinfo{title}{{Assessing Quantitative Results in Accretion Simulations: From Local to Global}}.
\newblock \emph{\bibinfo{journal}{\apj}} \textbf{\bibinfo{volume}{738}}, \bibinfo{pages}{84} (\bibinfo{year}{2011}).

\bibitem{sano2004}
\bibinfo{author}{{Sano}, T.}, \bibinfo{author}{{Inutsuka}, S.-i.}, \bibinfo{author}{{Turner}, N.~J.} \& \bibinfo{author}{{Stone}, J.~M.}
\newblock \bibinfo{title}{{Angular Momentum Transport by Magnetohydrodynamic Turbulence in Accretion Disks: Gas Pressure Dependence of the Saturation Level of the Magnetorotational Instability}}.
\newblock \emph{\bibinfo{journal}{\apj}} \textbf{\bibinfo{volume}{605}}, \bibinfo{pages}{321--339} (\bibinfo{year}{2004}).

\bibitem{abraham2018}
\bibinfo{author}{{Abraham}, Z.}
\newblock \bibinfo{title}{{Jet precession in binary black holes}}.
\newblock \emph{\bibinfo{journal}{\nas}} \textbf{\bibinfo{volume}{2}}, \bibinfo{pages}{443--444} (\bibinfo{year}{2018}).

\bibitem{britzen2018}
\bibinfo{author}{{Britzen}, S.} \emph{et~al.}
\newblock \bibinfo{title}{{OJ287: deciphering the `Rosetta stone of blazars$^{*}$'}}.
\newblock \emph{\bibinfo{journal}{\mnras}} \textbf{\bibinfo{volume}{478}}, \bibinfo{pages}{3199--3219} (\bibinfo{year}{2018}).

\bibitem{cruz2022}
\bibinfo{author}{{Cruz-Osorio}, A.} \emph{et~al.}
\newblock \bibinfo{title}{{State-of-the-art energetic and morphological modelling of the launching site of the M87 jet}}.
\newblock \emph{\bibinfo{journal}{\nas}} \textbf{\bibinfo{volume}{6}}, \bibinfo{pages}{103--108} (\bibinfo{year}{2022}).

\bibitem{mizuno2014}
\bibinfo{author}{{Mizuno}, Y.}, \bibinfo{author}{{Hardee}, P.~E.} \& \bibinfo{author}{{Nishikawa}, K.-I.}
\newblock \bibinfo{title}{{Spatial Growth of the Current-driven Instability in Relativistic Jets}}.
\newblock \emph{\bibinfo{journal}{\apj}} \textbf{\bibinfo{volume}{784}}, \bibinfo{pages}{167} (\bibinfo{year}{2014}).

\bibitem{singh2016}
\bibinfo{author}{{Singh}, C.~B.}, \bibinfo{author}{{Mizuno}, Y.} \& \bibinfo{author}{{de Gouveia Dal Pino}, E.~M.}
\newblock \bibinfo{title}{{Spatial Growth of Current-driven Instability in Relativistic Rotating Jets and the Search for Magnetic Reconnection}}.
\newblock \emph{\bibinfo{journal}{\apj}} \textbf{\bibinfo{volume}{824}}, \bibinfo{pages}{48} (\bibinfo{year}{2016}).

\end{thebibliography}

\begin{addendum}
\item[Data availability statement] 
The raw data can be downloaded from the EAVN Archive system (\url{https://radio.kasi.re.kr/arch/search.php}) and NRAO Archive Interface (\url{https://data.nrao.edu/portal/#/}). The calibrated data used in this paper are available from the corresponding author upon reasonable request due to the ongoing projects.

\item[Code availability statement] 
For data processing, we utilize public software, including AIPS for calibration (\url{http://www.aips.nrao.edu/index.shtml}), DIFMAP for imaging (\url{https://sites.astro.caltech.edu/~tjp/citvlb/}), and Python package {\tt EMCEE} for MCMC fitting (\url{https://pypi.org/project/emcee/}). The codes for the simulations in this paper are available from the corresponding author upon reasonable request due to the ongoing and follow-up projects.

\item[Acknowledgements] 
We thank James Moran and the other anonymous, reviewer(s) for their contributions to the peer review of this work. Wei Wang, Chiming Yim, Zhen Wang, Fangyuan Gu, Yi Feng, Masanori Nakamura, Shanshan Zhao, and Tsutomu Yanagida for the fruitful discussions and kind support. This project is funded by the China Postdoctoral Science Foundation (grant no. 2022M712084) and the Key Research Project of Zhejiang Lab no. 2021PE0AC03. Y.C. is supported by the Japanese Government (MEXT) Scholarship. This work is partially supported by the MEXT/JSPS KAKENHI (grant Nos. JP18H03721, JP19H01943, JP18KK0090, JP2101137, JP21H04488, JP22H00157, JP18K13594, JP19H01908, JP19H01906, JP18K03656, JP19KK0081). This work has been supported by the National Key R\&D Program of China (No. 2022YFA1603104), the Major Program of the National Natural Science Foundation of China (grant no. 11590780, 11590784), and Key Research Program of Frontier Sciences, CAS (Grant No. QYZDJ-SSW-SLH057). T.K. is supported in part by MEXT SPIRE, MEXT as "Priority Issue on post-K computer" (Elucidation of the Fundamental Laws and Evolution of the Universe) and as “Program for Promoting Researches on the Supercomputer Fugaku” (Toward a unified view of the universe: from large scale structures to planets, and Structure and Evolution of the Universe Unraveled by Fusion of Simulation and AI; Grant Number JPMXP1020230406), and JICFuS. The GRMHD simulations were carried out on the XC50 at the Center for Computational Astrophysics, National Astronomical Observatory of Japan. Y.M. is supported by the National Natural Science Foundation of China (No. 12273022) and the Shanghai pilot program of international scientists for basic research (No. 22JC1410600). J.Y.K. acknowledges the support from the National Research Foundation of Korea (No. 2022R1C1C1005255). S.T. acknowledges financial support from the National Research Foundation of Korea (NRF) grant 2022R1F1A1075115. This research was supported by the Korea Astronomy and Space Science Institute under the R\&D program supervised by the Ministry of Science and ICT. H.R. and B.W.S. acknowledge support from the KASI-Yonsei DRC program of the Korea Research Council of Fundamental Science and Technology (DRC-12-2-KASI). I.C. acknowledges financial support in part by the Consejería de Econom{\'i}a, Conocimiento, Empresas y Universidad of the Junta de Andaluc{\'i}a (grant P18-FR-1769), the Consejo Superior de Investigaciones Cient{\'i}ficas (grant 2019AEP112), and theSevero Ochoa grant CEX2021-001131-S funded by MCIN/AEI/ 10.13039/501100011033. R.-S.L. is supported by the Key Program of the National Natural Science Foundation of China (grant no. 11933007); the Key Research Program of Frontier Sciences, CAS (grant no. ZDBS-LY-SLH011); the Shanghai Pilot Program for Basic Research, Chinese Academy of Sciences, Shanghai Branch (JCYJ-SHFY-2022-013) and the Max Planck Partner Group of the MPG and the CAS. This work made use of the East Asian VLBI Network (EAVN), which is operated under cooperative agreement by National Astronomical Observatory of Japan (NAOJ), Korea Astronomy and Space Science Institute (KASI), Shanghai Astronomical Observatory (SHAO), Xinjiang Astronomical Observatory (XAO), Yunnan Astronomical Observatory (YNAO), National Astronomical Research Institute of Thailand (Public Organization) (NARIT), and National Geographic Information Institute (NGII), with the operational support by Ibaraki University (for the operation of Hitachi 32 m and Takahagi 32 m), Yamaguchi University (for the operation of Yamaguchi 32 m) and Kagoshima University (for the operation of VERA Iriki antenna). The Nanshan 26 m radio telescope (NSRT) is operated by the Urumqi Nanshan Astronomy and Deep Space Exploration Observation and Research Station of Xinjiang. The Sardinia Radio Telescope is funded by the Ministry of University and Research (MIUR), Italian Space Agency (ASI), and the Autonomous Region of Sardinia (RAS) and is operated as National Facility by the National Institute for Astrophysics (INAF). The Medicina[Noto] radio telescope is funded by the MIUR and is operated as National Facility by the INAF. The VLBA is an instrument of the National Radio Astronomy Observatory. The National Radio Astronomy Observatory is a facility of the National Science Foundation operated by Associated Universities, Inc.

\item[Author Contributions]
Y.C. leads the project. Y.C., K.H., H.R., K.Y., Jintao.Yu, J.P., W.J., and E.K. worked on the data calibration, image reconstruction, analysis, and interpretation of the results. T.K., M.K., W.L., Y.M., M.H., and Z.S. worked on the theoretical implications, simulations, and interpretation of the results. Y.C. wrote the original manuscript. J.-C.A., X.C., I.J., G.G., M.G., T.J., R.-S.L., K.N., J.O., K.O., S.S.-S., B.W.S., H.T., M.T., F.T., S.T., and K.W. contributed in the scientific discussions via EAVN AGN Science Working Group’s regular meetings. Kazunori Akiyama, T.A., K eiichi Aasada, S.B., D.B., L.C., Y.H., T.H., J.H., N.K., J.-Y.K., S.-S.L., J.W.L, J.A.L, G.M., A.Melis, A.Melnikov, C.M., S.-J.O, K.S., X.W., Y.Z., Z.C., J.-Y.H., D.-K.J., H.-R.K., J.-S.K., H.K., H.K., B.L., G.L., Xiaofei Li, Z.L., Q.L., Xiang Liu, C.-S.O., T.O., D.-G.R., J.W., N.W., S.W., B.X., H.Y., J.-H.Y., Y.Y., Jianping Yuan, H.Z., R.Z., and W.Z. worked on conducting observations, data correlation, and antenna maintenance. All authors contributed to the discussion of the results presented and commented on the manuscript.

\item[Author Information]
Reprints and permissions information is available at www.nature.com/reprints. The authors declare no competing interests. Readers are welcome to comment on the online version of this paper, which contains supplementary material and peer review reports available (\url{https://www.nature.com/articles/s41586-023-06479-6}). Correspondence and requests for materials should be addressed to Yuzhu Cui (yuzhu\_cui77@163.com).

\item[Competing interests] The authors declare no competing interests.

\end{addendum}

\end{document}